\documentclass[a4paper,fleqn]{cas-sc}

\usepackage[authoryear]{natbib}
\usepackage{graphicx,times}
\usepackage{subcaption}
\usepackage{amssymb,amsmath}
\usepackage{aas_macros}
\let\oldAA\AA
\renewcommand{\AA}{\text{\normalfont\oldAA}}
\usepackage{pdflscape}

\def\tsc#1{\csdef{#1}{\textsc{\lowercase{#1}}\xspace}}
\tsc{WGM}
\tsc{QE}

\begin{document}
\let\WriteBookmarks\relax
\def\floatpagepagefraction{1}
\def\textpagefraction{.001}

\shorttitle{Association of $\Gamma$ and optical parameters in AGNs}    

\shortauthors{D. Nour \& K. Sriram}  

\title [mode = title]{Dependency of Optical/UV parameters on X-ray spectral index in AGNs }  

\tnotemark[<tnote number>] 


%

\author[1,2]{D. Nour}

\cormark[1]


\ead{nour.abb.dergham@gmail.com}



\affiliation[1]{organization={Department of Astronomy, Osmania University},
            city={Hyderabad},
            postcode={500007}, 
            state={Telangana},
            country={India}}
            
\affiliation [2]{organization={Department of Physics, Faculty of Science, Al Baath University},
            city={Homs},
            country={Syria}}

\author[1]{K. Sriram}


\ead{ksriramou@osmania.ac.in}


\credit{}


\cortext[1]{Corresponding author}



\begin{abstract}
The study of Active Galactic Nuclei (AGNs) is vital in order to understand their respective nuclei bounded activity primarily triggered by the accretion disk and the associated corona. The hard X-ray emission characterised by the spectral index $\Gamma$ emitted by these sources in 2-10 keV band ionizes the nearby physical features like disk, BLR, NLR and provides a radiative configuration. However, based on previous studies, there is degeneracy in the evolution of $\Gamma$ across the redshift, wherein few studies display a systematic trend in the evolution and others rule out the systematic variation. In the present work, we study the evolution of $\Gamma$ across the redshift for a very large sample and perform further studies on the variation of $\Gamma$ with optical parameters using SDSS data viz. H$_{\beta}$ \& Mg II and their luminosities. Based on analysis we find that there is no change in $\Gamma$ across the redshift and it does not show any correlation with the optical parameters. This strongly suggests that the $\Gamma$ is influenced by the soft excess and/or reflection component and is affecting the disk and BLR region at the same strength across the redshift.
\end{abstract}




\maketitle

\section{Introduction}           
\label{sect:intro}
Active galactic nucleus (AGN) exhibits high activity due to accretion of material onto the super massive black hole (SMBH) which emits radiation at various luminosities with accretion rate L$_{\text{bol}}$/L$_{\text{Edd}}$ $>$ 10$^{-5}$ (\citealt{Yuan+etal+2014}; L$_{\text{Edd}}$ is the Eddington luminosity). AGNs can be divided into two categories known as type-1 and type-2 AGNs based on the width of the strong optical emission lines emanating primarily from two different regions (i.e. Broad line region (BLR) and narrow line region (NLR)). Type-1 AGNs are associated with emission from BLR that produces lines with FWHMs ranging from 1000-10000 km s$^{-1}$ (\citealt{Peterson+1997}) while type-2 AGNs exhibit FWHMs $\le$ 1000 km s$^{-1}$ arising from the orbiting clouds forming the NLR (\citealt{Blandford+etal+1990}). These orbiting clouds in random direction are photoionized by the high energy radiation exhibiting from the accretion disk close to the SMBH. The electron density of BLR ranges from 10$^{9}$ to 10$^{13}$ cm$^{-3}$, this region reprocesses almost 10\% of the disk radiation and produces spectral lines with an average energy around 10 eV pertaining to a photon density of $\sim$ 10$^{9}$ cm$^{-3}$ (\citealt{Abolmasov+etal+2017}). An important component of AGN structure is the dusty torus that primarily emits in infra-red (IR) wavelength, whose location is considered to be inside the kilo-parsec radius of NLR and outside of the BLR (e.g. \citealt{Antonucci+1993, Urry+etal+1995}). The type-2 AGN are obscured by a torus like structure with dust dominating features. Hence the BLRs are observed to be partially hidden, though they are detected in spectro-polarimetric studies (see, e.g. \citealt{Antonucci+1984, Antonucci+etal+1985, RamosAlmeida+etal+2016}).

The optical/UV radiation from the accretion disk is absorbed in the torus and re-emit in the near to mid-IR wavelengths in most of the AGNs. 
All the primary structures pertaining to BLR, NLR and dusty torus are primarily affected by the X-ray radiation produced in the inner region of the accretion disk. The X-ray radiation is thought to be generated in the Comptonization region or corona composed of hot electrons, which up-scatters the UV and optical photons via the inverse Compton scattering (\citealt{Haardt+etal+1991, Haardt+etal+1993, Liu+etal+2017}). The exact size of corona in AGNs is not properly understood but is estimated to be around 5-10 R$_{\text{g}}$ where R$_{\text{g}}$ = GM$_{\text{BH}}$/c$^{2}$ is the gravitational radius for a SMBH) based on the fast X-ray variability (e.g. \citealt{McHardy+etal+2005}) and short duration X-ray eclipses (e.g. \citealt{Risaliti+etal+2005, Risaliti+etal+2011}). Same size of corona was also estimated based on the micro-lensing studies (e.g. \citealt{Chartas+etal+2009}) and confirmed that almost half-light radius of corona $\sim$ 6 R$_{g}$ must exist in the inner region of accretion disk. A typical size of corona of about 3--10 R$_{g}$ is considered to be located above the disk, this also confirmed based on the reverberation studies of X-ray radiation reprocessed in the accretion disk close to the SMBH (e.g. \citealt{Fabian+etal+2009, Zoghbi+etal+2012, DeMarco+etal+2013, Kara+etal+2013}). But least has been understood how the corona radiative properties vary along with the accretion rate (\citealt{Xie+rtal+2017}) which has important consequences on the radiative heating that in-turn affects the radiative features of BLR and NLR. 

There is at least one main characteristic feature of hard X-ray radiation often parameterized by the spectral index ($\Gamma$) in the X-ray energy band 2-10 keV possibly originating in the corona. These X-rays affect the radiative and physical structure of the outer region of accretion as well as BLR and NLR. Moreover the effects of hard X-rays over the various components of H$_{\alpha}$, H$_{\beta}$, and MgII spectral lines are important as the later are often utilized to the constrain the mass of black holes as well as the size of BLR and NLR. The hard X-ray luminosity is strongly correlated with luminosity of optical lines which suggests that the corona has a tight control over the radiative properties of BLR and NLR regions (eg. \citealt{Jin+etal+2012a, Sriram+etal+2022}). Moreover the UV continuum emission and hard X-ray luminosity are found to exhibit a correlation (\citealt{Marchese+etal+2012}). The other important aspect of hard X-ray is the evolution of $\Gamma$ over a wide range of redshift. It has been found that $\Gamma$ does not vary significantly across redshift till z = 4. Both radio loud and radio quiet AGNs spectral index were not found to vary across a redshift till 2 (\citealt{Brinkmann+etal+1997,Yuan+etal+1998}). Moreover a flattened correlation between $\Gamma$ and redshift was noted by analyzing 473 quasar data using Sloan Digital Sky Survey (SDSS-DR5) along with XMM-Newton data(\citealt{Young+etal+2009}). Whereas \cite{Kelly+etal+2007} performed a systematic study of 174 AGNs using Chandra data and found $\Gamma$ to exhibit a marginal evolution across a redshift of 0.1--4.5. \cite{Shehata+etal+2021} also noted a flattened correlation i.e. $\Gamma$ is not varying across the redshift till 3--5.

H$_{\beta}$ and MgII widths and respective luminosities are crucial parameters to estimate the mass of the super massive black hole and exhibit the strong correlation to the hard X-rays emission arising from the inner region of accretion disk. At least two to three components are required to fit H$_{\beta}$ spectral profiles which explains the broad and narrow components of the line, whereas MgII line is often fitted with only two components. Generally it has been argued that the wings of the Mg II line are not associated with the production mechanism responsible for the core of the line and the physical origin causing the wings and the core is entirely different (\citealt{Jonic+etal+2016, Popovi+etal+2019}). It is important to study the dependency of these lines on the $\Gamma$ to understand how the respective lines are being affected due to hard X-ray emission.

\cite{Shehata+etal+2021} reported that $\Gamma$ is not varying along the redshift and its dependencies with various optical parameters are important to understand the radiative and physical configuration of the inner region of AGNs. 
The aim of the present study is to study the correlation between the hard X-ray luminosity (L$_{2-10}$ keV) \& $\Gamma$ (2-10 keV) with various optical and UV spectral components of H$_{\beta}$ and MgII spectral lines. Moreover, how $\Gamma$ is evolving with various optical spectral parameters is essential to understand the influence of corona over the BLR and NLR. The paper is organised in the following manner. Section 2 explain the data collection and reduction for a sample of 1512 AGNs spanning a redshift of z = 4. Before studying the dependency of $\Gamma$ and optical parameters in later sections, we show that the $\Gamma$ is not varying across the redshift in section 3.1. In section 4, we present the conclusion of the present study.

\section{Data collection and reduction}
\cite{Paris+etal+2018} catalog contains 526356 quasars confirmed quasars, 144 046 of them are new discoveries from SDSS-IV whose luminosities M$_{\text{i}}$ [z = 2] $<$ -20.5 and FWHM of at least one emission line is greater than 500 km s$^{-1}$ or show an absorption feature, the rest of the quasars are previously confirmed from SDSS-I, II, and III. 
\cite{Liu+etal+2016} X-ray catalog consists of 8554 AGNs form XMM-XXL north survey \citealt{Pierre+etal+2016}, the total area covered by this survey is of 50$^{\circ^{2}}$ and is split into two equal fields. 
We cross matched the optical and X-ray catalogs previously mentioned within a radius of 5 arcsec and by limiting the choice to Type I AGNs, this resulted in 1819 sources, one more criteria is to take sources for which hard X-ray luminosity and photon index are available. The final sample contains 1512 AGNs.

For the selected sample we downloaded optical spectra from the 14th data release of Sloan Digital Sky Survey (SDSS) database, we started analyzing theses spectra by applying the following corrections (check \citealt{Sriram+etal+2022} for more details)\\
1. Using the galactic extinction maps of \cite{Schlegel+etal+1998} and \cite{Fitzpatrick+etal+1999} parameters, we corrected the spectra for galactic reddening.\\
2. We used the relations $\lambda_ {\text{rest}}$ = $\lambda_ {\text{observed}}$ / (1+z) and f$_{\text{rest}}$ = f$_{\text{observed}}$ $\times$ (1+z) to correct wavelength and flux for redshift, respectively.\\
3. As we are interested to study the central part of the galaxy, then we subtracted the contribution of the host galaxy in the optical spectra based on the assumption of \cite{VandenBerk+etal+2006}, where each spectrum can be divided into a QSO part that accounts for the quasar part of the spectrum and another GAL part that accounts for the host galaxy contribution, the linear contribution of these two parts can reconstruct the AGN spectrum. We perform principal component analysis (PCA) technique using the first 10 QSO eigenspectra and the first 5 galaxy eigenspectra given in \cite{Yip+etal+2004a,Yip+etal+2004b} to reconstruct each spectrum.\\
4. For the QSO part of the spectum, we subtracted the continuum below H$\beta$ and MgII emission lines by interpolating the windows given in \cite{Kuraszkiewicz+etal+2002} on the sides of each emission .\\
5. Optical FeII lines exist in the range 4000-5500\AA\/ were fitted using the template given in (\citealt{Kova+etal+2010,Shapovalova+etal+2012}). The UV FeII that overlaps with MgII line was fitted following the method explained in \cite{kova+etal+2015}.\\
6. We adopted the two-component model (\citealt{Popovic+etal+2004,Bon+etal+2009}) to fit the emission line where emission form the broad line region can be modeled by two gaussians, one accounts for the core of emission line that originates from a region called as intermediate line region (ILR) and the other fits the wings of the line and is originating from a closer region to the SMBH and known as very broad line region (VBLR), one more Gaussian was used to fit the the narrow component of $H\beta$ line that originates from the narrow line region (NLR). MgII line was fitted after subtracting the UV FeII using only two Gaussian components viz. one for the core and another to fit the wings. Examples of the best fit of H$\beta$ and MgII lines are shown in Figures. 1-a and 1-b, respectively.
\begin{figure}[ht]
\begin{subfigure}{.5\textwidth}
  \centering
  \includegraphics[width=1.1\linewidth]{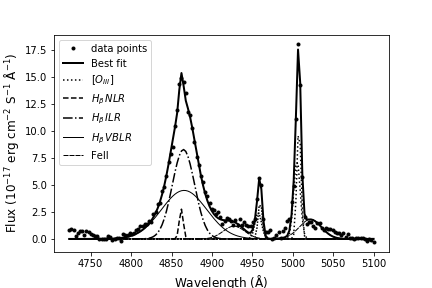}  
  \caption{H$\beta$ and [OIII] emission lines best fit.}
  \label{fig:sub-first}
\end{subfigure}%
\begin{subfigure}{.5\textwidth}
  \centering
  \includegraphics[width=1.1\linewidth]{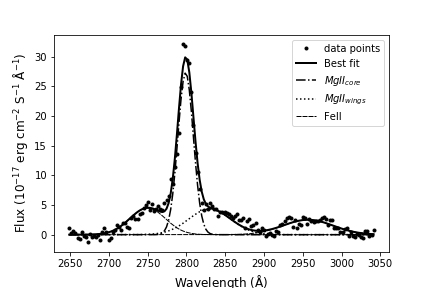}  
  \caption{MgII line best fit.}
  \label{fig:sub-second}
\end{subfigure}
\caption{Examples of best fit of (a) H$\beta$ and (b) and MgII lines of the source SDSS J021340.1-050149.1 }
\label{fig:fig}
\end{figure}
7. For each component of the fitted lines we calculated the luminosities and FWHMs , we also obtained the optical continuum luminosity at 5100$\AA$ and UV continuum luminosity at 3000 $\AA$.\\
8. To calculate the mass of the SMBH we used either L$_{5100\AA}$ \& FWHM of H$\beta$ or L$_{3000\AA}$ \& FWHM of MgII because due to the wavelength coverage of SDSS spectra, at high redshift the H$\beta$ line will shift out of the optical range. For this we used the following relations.
In case of H$\beta$ line (\citealt{Assef+etal+2011})
\begin{equation}
log\Big(\frac{{M}_{\text{BH}}}{M_{\odot}}\Big) =  0.89+0.5\,log\Bigg(\frac{\lambda L_{\lambda}(5100\,{\AA})}{10^{44}\, \text{erg\,s}^{-1}}\Bigg)+ 2 \,log     \Big(\frac{\text{FWHM}_{{H}\beta}^{{B}}}{\text{km\, s}^{-1}}\Big)
\end{equation}
In case of using MgII line (\citealt{Shen+etal+2011})
\begin{equation}
log\Big(\frac{{M}_{\text{BH}}}{M_{\odot}}\Big) =  0.74+0.62\,log\Bigg(\frac{\lambda L_{\lambda}(3000\,{\AA})}{10^{44}\, \text{erg\,s}^{-1}}\Bigg)+ 2 \,log     \Big(\frac{\text{MgII}}{\text{km\, s}^{-1}}\Big)
\end{equation}

The Bayesian X-ray Analysis package (\citealt{Buchner+etal+2014}) used to analyse the X-ray spectra combines both the Sherpa X-ray analysis (\citealt{Freeman+etal+2001}) and the Bayesian methodology which accurately propagates the uncertainties in the redshift as well as the X-ray observations, based on their description of the best model to fit a a large set of data, \cite{Liu+etal+2016} used to fit a large set of data is composed of three main components (BNTORUS + PEXMON + SCATTERING), each component fits part of the X-ray emission as follow: (i) BNTORUS (\citealt{Brightman+etal+2011}), it considered the X-ray source to be surrounded by a toroidal structure that causes photoelectric absorption and Compton scattering, hence it consists of a modified power law component to fit the power law continuum along with the absorption and scattering effects in the X-ray spectra. The inclination angle of this model is fixed at 85$^{\circ}$ whereas the opening angle is fixed at 45$^{\circ}$. 
(ii) PEXMON (\citealt{Nandra+etal+2007}), is used to describe the reflection of X-ray emission by the material surrounding the source, the inclination here is fixed at 60$^{\circ}$. (iii) SCATTERING, an unabsorbed power law is added to the model in order to account for the soft X-ray excess. From this study we adopted the photon index and the hard X-ray luminosity in the energy band (2-10keV).
Throughout this study we used the following  cosmological parameters for our calculations $\Omega_M$ = 0.3, $\Omega_\Lambda$ = 0.7, and $\Omega_k$ = 0, H$_0$ = 70 km s$^{-1}$ Mpc$^ {-1}$. 
All calculated and adopted parameters used in present study are listed in Table 1. 
\section{Discussion}
\subsection{Relation between X-ray photon index and redshift \& X-ray luminosity}
\cite{Shehata+etal+2021} studied the largest sample of 1280 quasars focusing on the evolution of $\Gamma$ with redshift. In their work they showed that the strong correlation usually reported between $\Gamma$ and z is caused by the soft excess or the reflection component and hence there is no evidence of spectral evolution in quasars. As discussed in section 2, the X-ray data adopted in our work were fitted with a multi-components model that account for the power-law, reflection, and scattering components of the X-ray spectrum, therefore in this work we investigated the relation between $\Gamma$ and redshift for one of the largest studied samples of 1512 AGNs.
We found a weak correlation between the two parameters ($\rho$ = -0.12, P = 2.15e-06, see Figure 2) that can be considered as nominal correlation. For more clarity we binned the data initially into 5 bins following the method explained in \cite{Scott+etal+2011} where for each bin, we calculated the weighted mean $\big<\Gamma\big>$ using the following equation
\begin{equation}
\big<\Gamma\big> = \sum_{i=0}^{n} P_i \times \Gamma_i\,;\,P_i=\frac{1/\sigma_i}{\sum_{i=0}^{n}(1/\sigma_i)^2}
\end{equation}
where n is the number of data points in each redshift bin and $\sigma_i$ is the error associated with each $\Gamma_i$. The errors on $\big<\Gamma\big>$ for each bin are considered as the standard errors on the mean and calculated using the equation
\begin{equation}
\alpha=\sigma/\sqrt{n}\,;\, \sigma = \sqrt{\frac{\sum(\Gamma_i-\big<\Gamma\big>)^2}{n-1}} 
\end{equation}
This binned data is shown in the second row of Fig. 2. We noted that there is almost no change in $\Gamma$ along the redshift, even when we tried to decrease / increase the bins sizes to include smaller / larger number of data points (Fig. 2, third and fourth rows), we did not find any change in the trend. Therefore this result confirms the finding of \cite{Shehata+etal+2021} that the reflection component might be the reason behind the null X-ray spectral index evolution of AGNs with redshift.
\label{sect:discussion}
\begin{figure} 
   \centering
   \includegraphics[width=15.0cm, angle=0]{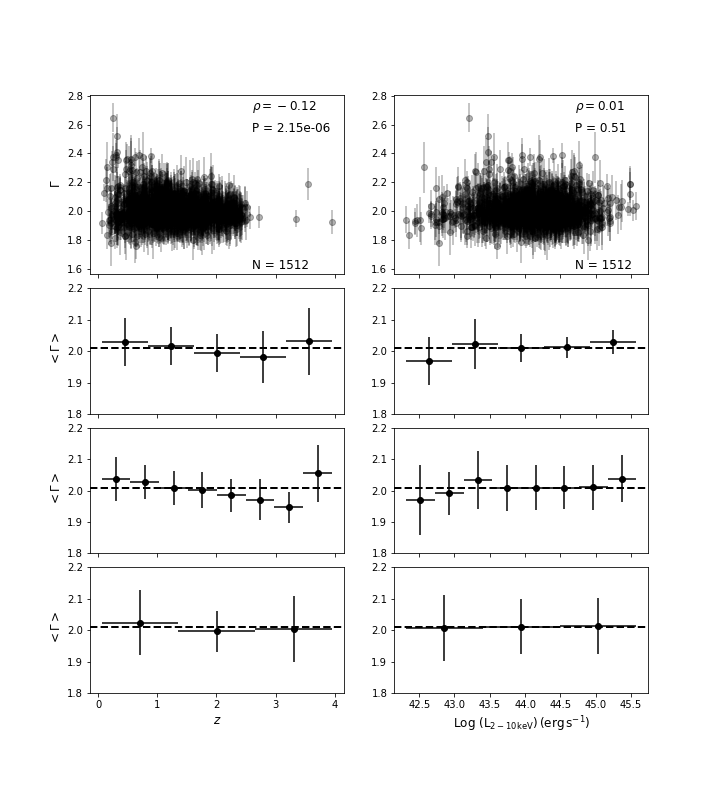}
   \caption{Upper panels: $\Gamma$ versus redshift (left) and Log (L$_{2-10\text{keV}}$) (right). The 2$^{\text{nd}}$, 3$^{\text{rd}}$ and 4$^{\text{th}}$ rows show the binned version of data in the upper panels. The dashed lines represent the weighted mean value of $\Gamma$ for the full sample $<\Gamma>$ = 2.01.} \label{Fig1}
   \end{figure}
We studied the relation between $\Gamma$ and L$_{2-10 \text{keV}}$ for the same sample as shown in the right panels of Fig. 2. In contrast to several previous findings, $\Gamma$ did not show any correlation with L$_{2-10 \text{keV}}$ ($\rho$ = 0.01, P = 0.51). Below we list briefly some studies that discussed this relation for samples of different sizes.\\
(i) \cite{Bianchi+etal+2009b} analyzed a sample of 156 X-ray unobscured, radio-quiet AGNs, with 0.002 $\leq$ z $\leq$ 4.520 from XMM-Newton Archive. 
The model adopted to fit the X-ray spectra contains two power-law components, one accounts for the hard emission, and the other for the soft excess. \cite{Bianchi+etal+2009a} stated that for this sample, the correlation between $\Gamma$ and L$_{2-10\text{keV}}$ is not significant.\\
(ii) In \cite{Jin+etal+2012a} a sample of 51 type I AGNs from XMM–Newton observations has been analyzed using a broad-band spectral energy distribution model that considers the emission from accretion disk along with hard X-ray emission. $\Gamma$ exhibits a significant anti-correlation with L$_{2-10 \text{keV}}$ ($\rho$ = -0.38).\\
(iii) \cite{Brightman+etal+2013} analyzed a sample from Chandra survey of 69 broad line, bright X-ray, radio-quiet AGNs. They reported a positive correlation between $\Gamma$ and L$_{2-10\text{keV}}$ where a single power-law model was used to fit the X-ray spectrum and no model was considered for the reflection component.\\
(iv) \cite{Trakhtenbrot+etal+2017} studied a low redshift sample of 228 AGN, and no correlation between $\Gamma$ and L$_{2-10\text{keV}}$ was observed when they unfolded X-ray spectra covering a broader energy band (BASS; 0.3-150 keV) and extracted the best fit spectral index through modelling, taking into account the soft excess, reflection, Fe line, warm absorber etc. wherever was found applicable.

By comparing the above four listed studies along with \cite{Shehata+etal+2021} and present study with the largest sample, we noted that, not only the relation between $\Gamma$ and redshift is affected by the model adopted to fit the X-ray spectra, but also the relation between $\Gamma$ and L$_{2-10\text{keV}}$ displays no correlation. Whereas, whenever a simple power-law model is used to fit the X-ray spectra, they confirm the existence of a strong correlation between $\Gamma$ and L$_{2-10\text{keV}}$. Otherwise, using a model that includes components to unfold the soft excess and/or reflection features, exhibit a very weak or no correlation between these two parameters. This present study confirms that $\Gamma$ is influenced with soft excess \& the reflection components and its null variation across the z signifies that both components are dominates AGN spectra at a similar strength.

\subsection{Relations between $\Gamma$ and Optical / UV continuum luminosities}

We discussed above how using an X-ray model that accounts for soft excess and reflection components to fit the X-ray spectrum as performed by \cite{Shehata+etal+2021} and the present work resulted in a non-evolving $\Gamma$ with redshift. Hence exploring its dependencies on other optical parameters are essential. We studied the relations between $\Gamma$ and optical/ UV parameters where the X-ray spectra are unfolded using a multi-component model as explained above. 
\begin{figure}[h!]
   \centering
   \includegraphics[width=15.0cm, angle=0]{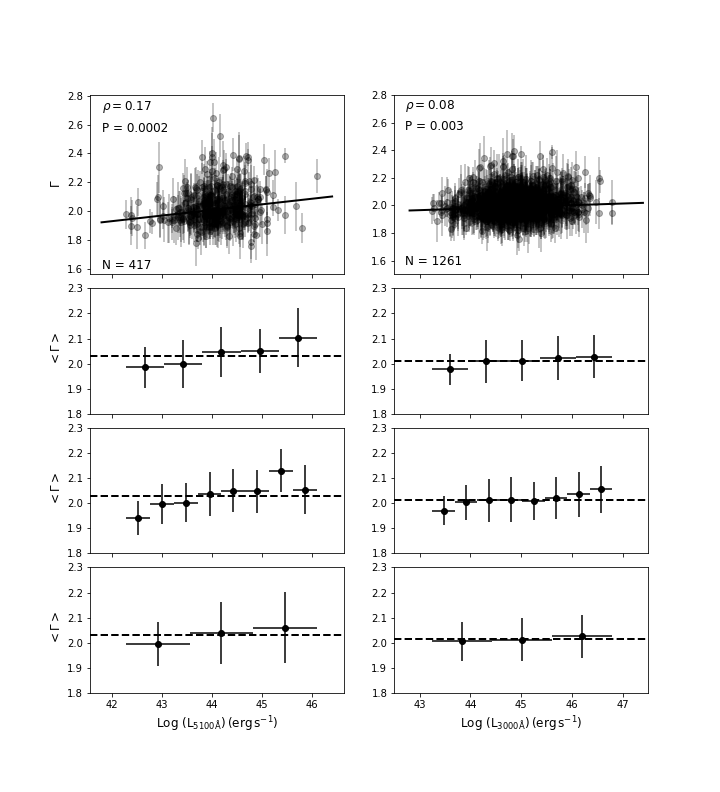}
   \caption{Upper panels: $\Gamma$ versus log ( L$_{5100\text{\AA}}$)  (left) and Log ($_{3000\text{\AA}}$) (right). The 2$^{\text{nd}}$, 3$^{\text{rd}}$ and 4$^{\text{th}}$ rows show the binned data of the upper panels. The dashed lines represent the weighted mean value of $\Gamma$ for the sample used in each upper panel.} 
   \label{Fig2}
   \end{figure}
According to the unified model of AGNs the accretion disk emits in optical and UV bands, therefore the continuum luminosities in these wavelengths are related to the accretion rate that usually affects the hardness of X-ray emission (\citealt{Sriram+etal+2022}).

\cite{Brightman+etal+2013} studied a sample of 32 AGNs and reported a strong correlation between $\Gamma$ and L$_{3000\text{\AA}}$ with a break at luminosities lower than L$_{3000\text{\AA}}$ = 10$^{44}$ erg s$^{-1}$. In the present sample, $\Gamma$ correlates weakly with L$_{5100\text{\AA}}$ and the linear best fit of the relation shows a very flat slope (slope = 0.03) (Fig. 2, left panels). No correlation was seen with L$_{3000\text{\AA}}$ as the correlation coefficient is $\sim$ 0, the slope of linear best fit displays a relatively shallower slope than that observed with  L$_{5100\text{\AA}}$ case (slope = 0.01) (Fig. 3, right panels). We did not find any indication of a break point in the relation between $\Gamma$ and L$_{3000\text{\AA}}$ even after binning the data. We noted that the weighted mean values of $\Gamma$ in each bin did not display a significant deviation from the weighted mean value of the sample studied, and this trend is not affected by the size of the bin, where using smaller or larger bins as shown in Fig. 3 gives the same result. This might be due to the fact that the sample is large (1261 AGNs for  z $<$ 4) and 83, among the sample have luminosities less than 10$^{44}$ erg s$^{-1}$ compared to 15 AGNs in \cite{Brightman+etal+2013}. However, our result is well in agreement with the studies of \cite{Young+etal+2009} where a sample of about 500 quasars was considered and other smaller samples (e.g.   \citealt{Page+etal+2004,Shemmer+etal+2005, Kelly+etal+2007}). In these studies no significant correlation was found between UV luminosity at 2500 $\text{\AA}$ and $\Gamma$.

\subsection{Hard X-rays and optical/UV emission lines parameters}
\subsubsection{$\Gamma$ versus FWHM of H$\beta$ and MgII emission lines}
The relations between the hard / soft X-ray photon index and FWHM of H$\beta$ have been investigated in several studies (e.g. \citealt{Reeves+etal+2000,Porquet+etal+2004,Piconcelli+etal+2005,Ojha+etal+2020}),  they reported a significant anti-correlation between the photon index and H$\beta$. However \cite{Sriram+etal+2022} stated that there exists a break point in this relation confirming the previous study by \cite{Jin+atal+2012} where the anti-correlation observed before the break point turns out to be a positive correlation at higher FWHMs values. A similar trend was also observed in the relations between $\Gamma$ and FWHM of H$\alpha$ \& M$_{\text{BH}}$. We tested this relation for the selected sample in this present study and we did not find $\Gamma$ to correlate with FWHM of H$\beta$ (BLR), similarly for FWHMs of ILR and VBLR components (Fig. 4). We also investigated the relation in case of FWHM of MgII, neither the core nor the wing components show a significant correlation with $\Gamma$ (Fig. 5). This result is inconsistent with the study of \cite{Brightman+etal+2013} that found a significant anti-correlation between $\Gamma$ and FWHM$_{\text{MgII}}$. The FWHM component of H$\beta$ and MgII core is completely Virialized and solely controlled by the mass of the black hole whereas, we argue that the $\Gamma$ is affected by many radiative processes viz. soft excess/reflection component \& coronal properties and their association to the mass of black hole is still not understood and hence no strong correlation is seen.
\begin{figure}[h!] 
   \centering
   \includegraphics[width=15.0cm, angle=0]{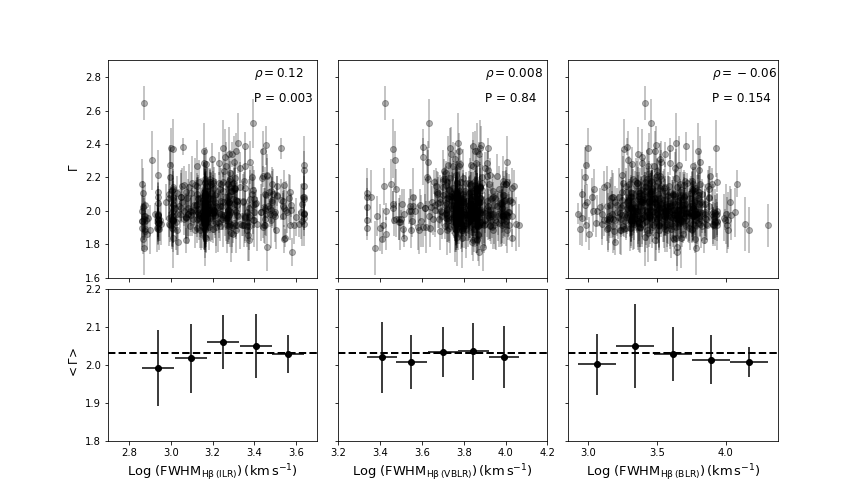}
   \caption{Upper panels represent the relations between $\Gamma$ and FWHM$_{H\beta}$ line components (i.e. ILR, VBLR, and BLR from left to right panel, respectively).Lower panels show the binned data of the upper panels. The dashed lines represent the weighted mean value of $\Gamma$ for the sample used in the upper panels.
   } 
   \label{Fig2}
   \end{figure}

\begin{figure}[h!] 
   \centering
   \includegraphics[width=15.0cm, angle=0]{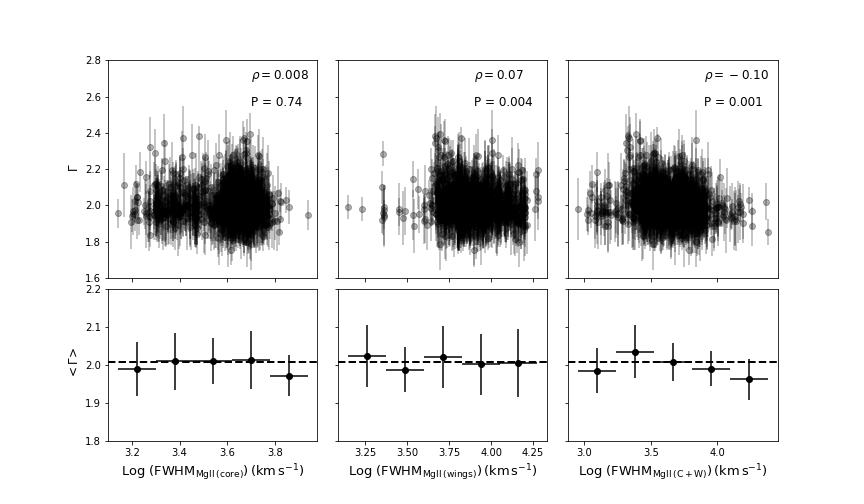}
   \caption{Upper panels represent the relations between $\Gamma$ and FWHM$_{\text{MgII}}$ line components (i.e. core, wing, and core+wing, from left to right panel, respectively). Lower panels show the binned data of the upper panels. The dashed lines represent the weighted mean value of $\Gamma$ for the sample used in the upper panels.
   } 
   \label{Fig2}
   \end{figure}
 
\subsubsection{$\Gamma$ versus Luminosities of H$\beta$ and MgII emission lines}
Hard X-ray emission in 2-10 keV energy band is considered to be produced by Comptonization process of hot electrons in the corona above the accretion disk (\citealt{RamosAlmeida+etal+2017}). However the spectral index, $\Gamma$, does not explicitly constrain the coronal properties because it is also affected by the soft or/and reflection component due to redshift (\citealt{Trakhtenbrot+etal+2017,Shehata+etal+2021}). On the other hand, the broad component of H$\beta$ and the core component of MgII are originating from the broad line region where the emitting region of MgII line is located at larger radius (\citealt{Popovi+etal+2019}), unlike the wing component that is caused by the inflow/outflow in the BLR (\citealt{savic+etal+2020}). We found a weak positive correlation between $\Gamma$ and luminosities of broad H$\beta$ line components (ILR, VBLR, and BLR) (Fig. 6). This indicates that the soft excess/reflection components oversee the radiative evolution of the BLR. In the case of MgII line, only the core component luminosity displays a very weak correlation with a scatter in the relation (Fig. 6). 
\begin{figure}[h!] 
   \centering
   \includegraphics[width=15.0cm, angle=0]{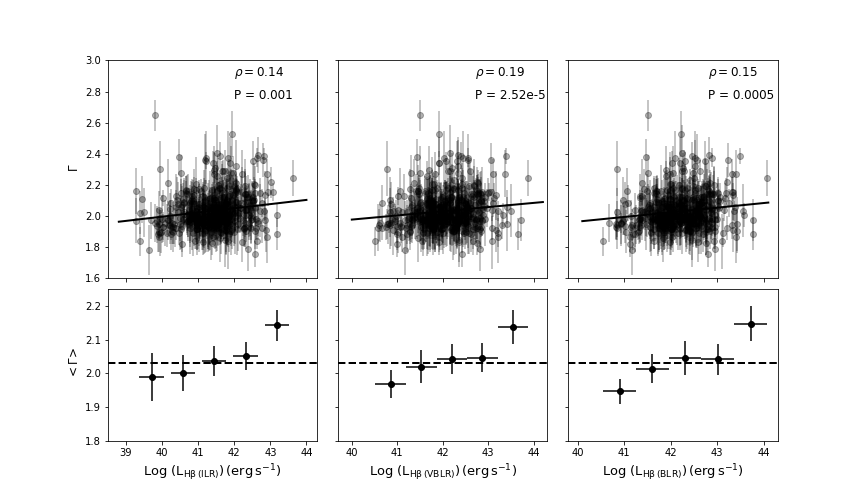}
   \caption{Upper panels represent the relations between $\Gamma$ and L$_{H\beta}$ line components (i.e. ILR, VBLR, and BLR from left to right panel, respectively).Lower panels show the binned data of the upper panels. The dashed lines represent the weighted mean value of $\Gamma$ for the sample used in the upper panels.
   } 
   \label{Fig2}
   \end{figure}
\begin{figure}[h!] 
   \centering
   \includegraphics[width=15.0cm, angle=0]{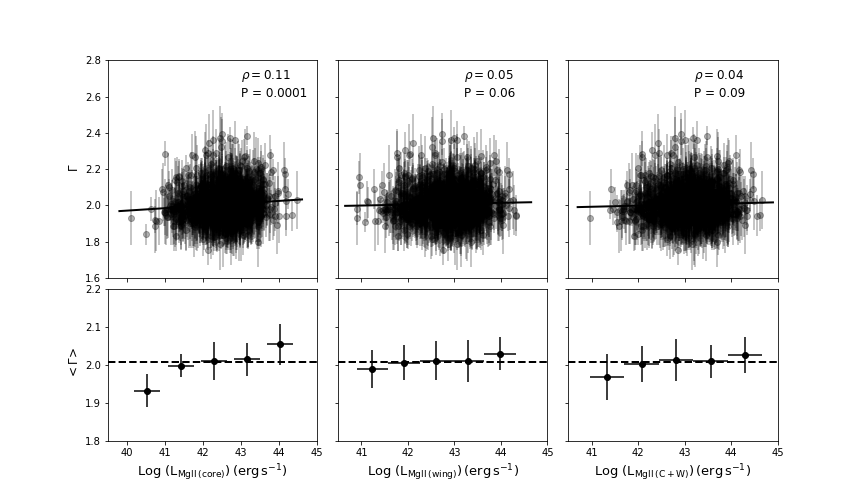}
   \caption{Upper panels represent the relations between $\Gamma$ and L$_{\text{MgII}}$ line components (i.e. core, wing, and core+wing, from left to right panel, respectively). Lower panels show the binned data of the upper panels. The dashed lines represent the weighted mean value of $\Gamma$ for the sample used in the upper panels.
   } 
   \label{Fig2}
   \end{figure}
Although using multi-component model to fit the X-ray spectrum caused results that differs from previous studies between $\Gamma$ and optical/UV parameters, but the strong correlation exists between L$_{2-10 \text{keV}}$ and luminosities of H$\beta$ as well as MgII components is not affected as shown in Fig. 8. We can say that the L$_{2-10 \text{keV}}$ is generated due to the varying combination of soft excess and/or reflection and coronal emissions, is strongly affecting the BLR region luminosities emitting H$\beta$ and MgII lines. The best fit relations between L$_{2-10 \text{keV}}$ and luminosities of BLR component (ILR+VBLR) of H$\beta$ \& MgII (core+wing) are given in equations 5 and 6, respectively.

\begin{figure}[h!] 
   \centering
   \includegraphics[width=15.0cm, angle=0]{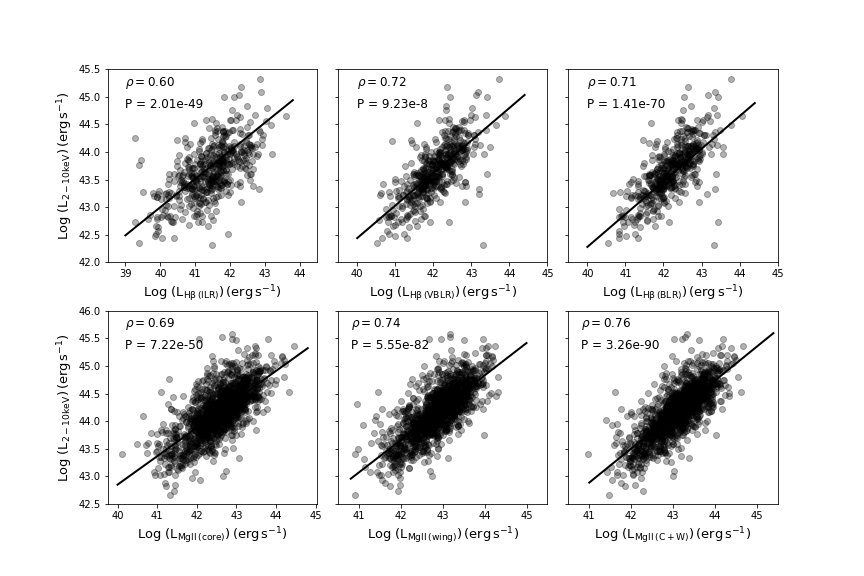}
   \caption{Upper panels show the correlations between $\Gamma$ and L$_{\text{H}\beta}$ line components. Lower panels shows the correlations between $\Gamma$ and L$_{\text{MgII}}$ line components.
   } 
   \label{Fig2}
   \end{figure}
   
\begin{align}
\log(L_{2-10\,\text{keV}})&=(0.59\pm0.02) \log(L_{\text{H}\beta^{\text{B}}})+(18.60\pm1.13)\\
\log(L_{2-10\,\text{keV}})&=(0.61\pm0.01) \log(L_{\text{MgII}^{\text{c+w}}})+(17.67\pm0.61)
\end{align}
\subsection{Dependency of $\Gamma$ on Eddington ratio}
Studying Eddington ratio is considered to be a pivotal parameter to understand the evolution of SMBHs. It can be obtained by estimating the ratio between bolometric and Eddington luminosities ($\lambda_{\text{Edd}}$ = L$_{\text{bol}}$/ L$_{\text{Edd}}$). We determined the bolometric luminosity as L$_{\text{bol}}$ = 20 $\times$ L$_{2-10 \text{keV}}$ (erg s$^{-1}$) (\citealt{Vasudevan+etal+2009}) and to estimate the Eddington luminosity we used the relation L$_{\text{Edd}}=1.3\times 10^{38} \text{M}_{\text{BH}}/\text{M}_\odot\, (\text{erg}\, \text{s}^{-1})$, where the masses of SMBHs were estimated using FWHMs of broad components of H$\beta$ line for AGNs at z $<$ 1.04 (equation 1), for sources at higher z, we used FWHMs of MgII to estimate the masses of SMBH (since H$\beta$ line will be shifted out of SDSS spectral range , see equation 2). We investigated the relation between $\Gamma$ and $\lambda_{\text{Edd}}$ for the full sample and statistically we found a weak positive correlation ($\rho$ = 0.15, P = 1.32e-08; Fig. 6-a) with a very flat slope $\sim$0.02. We divided the sample based on the emission line which was used to calculate the black hole mass (i.e H$\beta$ or MgII). By considering only H$\beta$ line, we found a stronger correlation ($\rho$ = 0.24, P = 9.48e-08; Fig. 6-b) with a slope that is slightly higher than the one obtained from the full sample (slope = 0.04). Eddington ratios obtained using MgII lines did not show any correlation with $\Gamma$ and the slope was found to be  $\sim$ 0.007. Even after we binned the data as shown in lower panels of Fig. 6, we observed no significant change in $\Gamma$ with Eddington ratio. This result is in agreement with a recent study by \cite{Kamraj+etal+2022}, where a sample of Swift/BAT selected Seyfert 1 AGNs modeled with a reflection model that accounts precisely for the radiation reprocessed from the corona, similar to our result, their slope was found to be $\sim$0.03–0.06 with $\rho$ $<$ 0.25. They conclude that there is no strong evidence for correlation between $\Gamma$ and $\lambda_{\text{Edd}}$.
However, these results stands in contrast with previous studies that reported a strong correlation between $\Gamma$ and $\lambda_{\text{Edd}}$ when $\lambda_{\text{Edd}}$ is greater than 10$^{-3}$ (e.g. \citealt{Shemmer+etal+2006, Shemmer+etal+2008, Zhou+etal+2010, Fanali+etal+2013}) with slope $\sim$ 0.3. The common feature among these studies is that the model used to fit the X-ray spectra contains a single power-law component. When \cite{Winter+etal+2012} analyzed the X-ray spectra of 51 Seyfert sources using a multi components model that includes a cutoff power-law model, a black-body model to account for the soft excess as well as a reflection component, they noted a flatter slope $\sim$ 0.23 of the relation. \cite{Ricci+etal+2013} reported also a flat slope $\sim$ 0.12 for a sample of 36 AGNs.
Over all we concluded that the relation between $\Gamma$ and $\lambda_{\text{Edd}}$ is not robust enough to be used as an estimator of Eddington luminosity or M$_{BH}$, because as discussed above the slope of this relation shows different values for each selected sample, depending on the model used to fit their X-ray spectrum. The $\Gamma$ is affected by both reflection and/or soft excess which is similar throughout the redshift, at least in our present sample. The reflection component fraction is considered to be higher for large coronal heights above the disk plane. \cite{DeMarco+etal+2013} found that the reverberation lags caused by the reflection component have a dependency on the mass of the black hole i.e. for higher mass of a BH, the observed lags are larger. If the Eddington ratio is high, it indicates toward a higher mass accretion rate which would cool down the corona, decreasing the reflection fraction. Since $\Gamma$ is influenced by the reflection component/soft excess, hence does not exhibit any variation or trend with other observed parameters in the present study. Future missions viz. XRISM (\citealt{XRISM+etal+2022}) and Athena-X (\citealt{Barret+etal+2018}) would be able to robustly constrain the X-ray emission in the band $\sim$ 0.3-12 keV which shall ascertain the studied relationships.

\begin{figure} 
   \centering
   \includegraphics[width=15.0cm, angle=0]{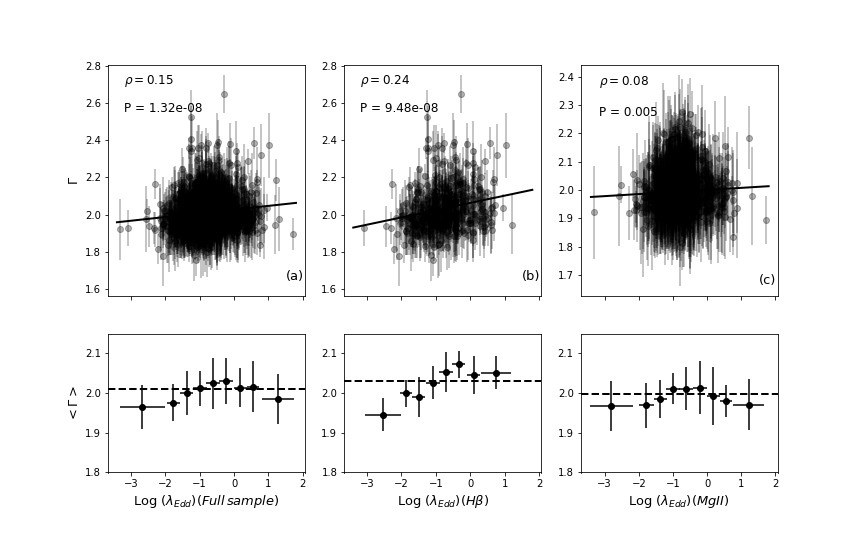}
   \caption{Upper panels show the relation between $\Gamma$ and $\lambda_{Edd}$ where $\lambda_{Edd}$ is calculated using  masses obtained from: a) FWHMs of both MgII and H$\beta$ lines, b) FWHMs of only H$\beta$ line, c) FWHMs of only MgII line. The solid lines show the linear best fit line.
   Lower panels show the binned version of upper panels. The dashed lines represent the weighted mean for the sample used in each panel.} 
   \label{Fig2}
   \end{figure}   

\section{Conclusions}
\label{sect:conclusion}

For a large sample of Type I AGNs (1512), we analyzed their SDSS spectra and studied the relations between the various optical/UV and hard X-ray parameters. The X-ray spectra were fitted using a multi components model to account for different emissions (\citealt{Liu+etal+2016}). Based on our results, we argue that the $\Gamma$ in 2-10 keV is not a sole indicator of corona and it is affected by the reflection and/or soft excess features. Hence the estimation of BH mass and any other parameter based on $\Gamma$ should be considered carefully. Below are main results of the present work
\begin{enumerate}
 \item We confirm the non evolving $\Gamma$ with redshift, and the evolution might be due to reflection or soft excess component.

 \item For this X-ray fitting model, the effect of $\Gamma$ on the kinematics of the Optical/UV emitting region is not observed anymore.

 \item The relation between $\Gamma$ and $\lambda_{Edd}$ is not robust to be used to estimate M$_{\text{BH}}$ and Eddington ratio.

 \item Except the correlation between hard X-ray luminosity and L$\text{H}\beta$ \& MgII, adopting a high sensitive X-ray spectral model to fit the X-ray spectrum causes a change in all other well known correlations as none of the relations studied in this work show a strong correlation.
\end{enumerate}

\section*{Acknowledgements}
We acknowledge the Referee for the helpful suggestions. D. Nour acknowledges Al Baath University, Syria for the financial support.
KS acknowledges the financial support from the Core Research Grant (CRG) scheme under SERB, Government of India.\\
Funding for SDSS-III has been provided by the Alfred P. Sloan Foundation, the Participating Institutions, the National Science Foundation, and the U.S. Department of Energy Office of Science. The SDSS-III web site is http://www.sdss3.org/.

SDSS-III is managed by the Astrophysical Research Consortium for the Participating Institutions of the SDSS-III Collaboration including the University of Arizona, the Brazilian Participation Group, Brookhaven National Laboratory, Carnegie Mellon University, University of Florida, the French Participation Group, the German Participation Group, Harvard University, the Instituto de Astrofisica de Canarias, the Michigan State/Notre Dame/JINA Participation Group, Johns Hopkins University, Lawrence Berkeley National Laboratory, Max Planck Institute for Astrophysics, Max Planck Institute for Extraterrestrial Physics, New Mexico State University, New York University, Ohio State University, Pennsylvania State University, University of Portsmouth, Princeton University, the Spanish Participation Group, University of Tokyo, University of Utah, Vanderbilt University, University of Virginia, University of Washington, and Yale University.

\begin{landscape}
\begin{table}
\caption{Spectral parameters used in present study for the first 50 objects of the full sample of 1512 AGNs. Column 3: redshift; column 4: continuum luminosity at 3000\AA; column 5: continuum luminosity at 5100\AA; columns 6,7 and 8: luminosities of H$\beta$ line components; columns 9, 10 and 11: luminosities of MgII line components, columns 12, 13 and 14: FWHMs of H$\beta$ components; columns 15, 16 and 17: FWHMs of MgII components; column 18: mass of the black hole estimated using equations (1) and (2); columns 19 and 20 X-ray luminosity and photon index adopted from \cite{Liu+etal+2016}. column 21: Eddington ratio}

\label{tab:landscape}
\setlength\tabcolsep{3pt}
\setlength{\cmidrulekern}{0.5em}
    \centering 

\scriptsize
\begin{tabular}{l|l|c|c|c|ccc|ccc|ccc|ccc|c|c|c|c}
\hline
 & & & & & & & & & & & & & & & & & & & \\

& &  &Log(L$_{3000\text{\AA}}$)&Log(L$_{5100\text{\AA}}$)&\multicolumn{3}{c|}{ Log(L$_{\text{H}\beta}$) (erg s$^{-1}$)}&\multicolumn{3}{c|}{Log(L MgII) (erg s$^{-1}$)}&\multicolumn{3}{c|}{Log(FWHM H$\beta$) (km s$^{-1}$)}&\multicolumn{3}{c|}{Log(FWHM MgII)(km s$^{-1}$)}&Log(M$_{BH}$)&Log(L$_{\text{X}}$)& & \\
& & & & & & & & & & & & & & & & & & & & \\
 
 \cline{6-8} \cline{9-11} \cline{12-14} \cline{15-17} 
ID &SDSS name& z&(erg s$^{-1}$) &(erg s$^{-1}$) & & & & & & & & & & & & &($\text{M}_{\odot}$) &(erg s$^{-1}$) &$\Gamma$ &$\lambda_{Edd}$ \\

& &  & & & 
  ILR & VBLR & BLR  & ILR & VBLR & BLR & core & wings & c+w& core & wings & c+w &  & & & \\
 (1) & (2) & (3) & (4) & (5) & (6) & (7) & (8) & (9) & (10) & (11) & (12) & (13) & (14) & (15) & (16) & (17) & (18)& (19)&(20)&(21)\\

\hline
1 & 020028.81-051618.2 & 1.530 & - & - & - & - & - & 42.230 & 42.887 & 43.299 & - & - & - & 3.776 & 3.798 & 3.841 & 8.787 & 44.312 & 1.980 & -1.314 \\
2 & 020108.13-051340.5 & 1.626 & 45.059 & - & - & - & - & 42.906 & 43.265 & 43.422 & - & - & - & 3.627 & 3.809 & 3.547 & 8.347 & 44.290 & 1.989 & -0.896 \\
3 & 020125.09-051120.1 & 0.824 & 44.162 & 43.752 & 41.784 & 41.759 & 42.073 & 42.235 & 42.402 & 42.627 & 3.399 & 3.858 & 3.464 & 3.671 & 3.733 & 3.417 & 7.267 & 43.670 & 2.072 & -0.436 \\
4 & 020129.32-051128.1 & 0.793 & - & - & 41.141 & 41.979 & 42.214 & 41.231 & 41.730 & 42.130 & 3.104 & 3.680 & 3.386 & 3.598 & 3.794 & 3.333 & 7.152 & 43.786 & 1.953 & -0.205 \\
5 & 020135.51-065110.6 & 2.408 & - & - & - & - & - & 42.637 & 43.284 & 43.633 & - & - & - & 3.221 & 3.899 & 3.128 & 7.674 & 44.473 & 2.046 & -0.039 \\
6 & 020139.25-050118.7 & 0.231 & - & 43.907 & 40.796 & 41.859 & 41.895 & - & - & - & 3.096 & 3.754 & 3.357 & - & - & - & 8.492 & 43.847 & 1.978 & -1.485 \\
7 & 020143.49-050913.5 & 1.279 & 44.382 & - & - & - & - & 42.459 & 42.762 & 42.938 & - & - & - & 3.652 & 3.803 & 3.524 & 7.197 & 44.099 & 1.998 & 0.062 \\
8 & 020153.27-050840.2 & 1.603 & 45.435 & - & - & - & - & 43.191 & 43.213 & 43.503 & - & - & - & 3.628 & 4.183 & 3.864 & 8.650 & 44.767 & 2.005 & -0.722 \\
9 & 020156.86-052109.8 & 1.621 & 46.111 & - & - & - & - & 43.344 & 44.053 & 44.131 & - & - & - & 3.708 & 3.802 & 3.722 & 8.982 & 44.365 & 1.956 & -1.456 \\
10 & 020202.44-042819.5 & 0.723 & 43.923 & 43.732 & 41.311 & 42.056 & 42.128 & 42.099 & 41.648 & 41.648 & 3.010 & 3.899 & 3.197 & 3.293 & 3.533 & 3.204 & 8.323 & 43.715 & 1.946 & -1.447 \\
11 & 020229.62-043128.1 & 1.686 & 44.850 & - & - & - & - & 42.856 & 42.789 & 43.125 & - & - & - & 3.689 & 4.190 & 3.851 & 8.877 & 44.386 & 2.011 & -1.329 \\
12 & 020235.80-064358.3 & 1.116 & 45.097 & - & - & - & - & 43.086 & 43.134 & 43.412 & - & - & - & 3.657 & 4.129 & 3.821 & 8.791 & 44.524 & 2.040 & -1.106 \\
13 & 020237.34-051115.2 & 0.434 & - & 43.984 & 40.313 & 41.560 & 41.584 & - & - & - & 2.946 & 3.758 & 3.136 & - & - & - & 7.962 & 43.251 & 1.892 & -1.551 \\
14 & 020241.79-050812.5 & 2.146 & 45.058 & - & - & - & - & 42.122 & 42.733 & 43.135 & - & - & - & 3.673 & 3.968 & 3.854 & 8.378 & 44.719 & 1.956 & -0.499 \\
15 & 020246.50-051255.2 & 3.958 & - & - & - & - & - & 42.054 & 42.621 & 43.003 & - & - & - & 3.820 & 3.969 & 4.059 & 8.492 & 44.083 & 1.925 & -1.248 \\
16 & 020246.57-050927.6 & 0.861 & 44.517 & 44.335 & 41.660 & 42.290 & 42.381 & 42.300 & 42.732 & 42.869 & 3.169 & 3.713 & 3.452 & 3.661 & 3.824 & 3.595 & 8.127 & 44.133 & 2.047 & -0.834 \\
17 & 020256.70-064805.6 & 2.039 & 45.343 & - & - & - & - & 42.920 & 42.407 & 43.036 & - & - & - & 3.308 & 3.842 & 3.451 & 8.356 & 44.693 & 2.036 & -0.502 \\
18 & 020256.82-042247.2 & 0.303 & - & 43.955 & 41.641 & 41.660 & 41.952 & - & - & - & 3.380 & 3.798 & 3.461 & - & - & - & 6.600 & 43.198 & 2.073 & -0.241 \\
19 & 020257.39-051225.4 & 0.512 & - & - & 40.603 & 41.388 & 41.597 & 41.161 & 41.683 & 42.050 & 3.242 & 3.833 & 3.901 & 3.722 & 3.929 & 3.570 & 8.201 & 43.047 & 1.981 & -1.994 \\
20 & 020301.97-051542.6 & 1.633 & 44.520 & - & - & - & - & 42.475 & 42.775 & 42.951 & - & - & - & 3.604 & 3.898 & 3.636 & 7.674 & 44.239 & 1.970 & -0.274 \\
21 & 020302.22-051814.2 & 1.513 & 45.018 & - & - & - & - & 42.953 & 42.835 & 43.199 & - & - & - & 3.468 & 4.199 & 3.895 & 8.865 & 44.460 & 1.963 & -1.244 \\
22 & 020306.44-045935.7 & 0.999 & 45.192 & 44.861 & 42.168 & 42.966 & 43.031 & 43.145 & 43.545 & 43.690 & 3.249 & 3.780 & 3.615 & 3.697 & 3.960 & 3.735 & 8.413 & 44.584 & 2.051 & -0.668 \\
23 & 020307.01-041649.5 & 1.049 & 44.365 & - & - & - & - & 42.348 & 42.549 & 42.761 & - & - & - & 3.500 & 3.942 & 3.664 & 8.151 & 43.908 & 1.997 & -1.082 \\
24 & 020309.07-043709.7 & 0.602 & 43.784 & 42.903 & 40.883 & 40.674 & 41.092 & 41.170 & 41.721 & 41.829 & 3.012 & 3.338 & 3.029 & 3.300 & 3.775 & 3.158 & 6.863 & 43.423 & 2.084 & -0.279 \\
25 & 020309.46-064133.3 & 1.407 & 45.646 & - & - & - & - & 43.200 & 43.509 & 43.682 & - & - & - & 3.645 & 3.803 & 3.506 & 8.094 & 45.209 & 1.920 & 0.276 \\
26 & 020309.71-050750.3 & 1.805 & 45.006 & - & - & - & - & 42.048 & 42.906 & 42.962 & - & - & - & 3.602 & 3.803 & 3.812 & 8.477 & 44.618 & 2.005 & -0.698 \\
27 & 020311.48-050500.7 & 1.250 & 45.108 & - & - & - & - & 42.380 & 43.051 & 43.135 & - & - & - & 3.693 & 3.825 & 3.797 & 7.454 & 44.519 & 1.995 & 0.226 \\
28 & 020313.15-051057.6 & 0.514 & - & - & 40.165 & 40.907 & 40.761 & 40.935 & 41.476 & 41.798 & 3.186 & 3.836 & 3.692 & 3.574 & 3.829 & 3.510 & 7.714 & 43.034 & 2.073 & -1.519 \\
29 & 020320.25-045917.3 & 1.028 & - & - & 40.752 & 41.552 & 41.768 & 41.707 & 42.311 & 42.713 & 3.188 & 3.839 & 3.702 & 3.752 & 3.939 & 3.710 & 7.854 & 44.090 & 2.039 & -0.602 \\
30 & 020320.47-050933.8 & 1.353 & 44.338 & - & - & - & - & 42.388 & 42.470 & 42.732 & - & - & - & 3.767 & 3.900 & 3.568 & 7.469 & 44.044 & 1.940 & -0.263 \\
31 & 020324.14-070940.8 & 1.118 & 45.478 & - & - & - & - & 43.000 & 43.516 & 43.632 & - & - & - & 3.680 & 3.830 & 3.591 & 8.453 & 45.463 & 2.014 & 0.171 \\
32 & 020326.22-051020.5 & 0.891 & 45.598 & 45.458 & 42.829 & 43.382 & 43.489 & 43.161 & 43.216 & 43.490 & 3.059 & 3.606 & 3.284 & 3.480 & 3.667 & 3.326 & 8.308 & 44.392 & 2.385 & -0.755 \\
33 & 020327.04-045853.4 & 0.972 & 44.987 & 44.221 & 42.568 & 42.812 & 43.008 & 42.635 & 43.001 & 43.156 & 3.285 & 3.780 & 3.413 & 3.652 & 3.705 & 3.430 & 7.766 & 44.540 & 2.144 & -0.065 \\
34 & 020327.73-064142.4 & 1.641 & 44.990 & - & - & - & - & 42.918 & 42.995 & 43.259 & - & - & - & 3.738 & 4.140 & 3.836 & 8.268 & 45.061 & 1.951 & -0.046 \\
35 & 020328.65-045737.8 & 0.765 & 44.657 & 44.288 & 42.285 & 42.574 & 42.754 & 42.727 & 42.931 & 43.142 & 3.282 & 3.700 & 3.427 & 3.653 & 3.685 & 3.370 & 6.900 & 44.412 & 2.190 & 0.673 \\
36 & 020330.88-065536.4 & 0.856 & 44.829 & 44.787 & 42.010 & 42.399 & 42.547 & 42.294 & 42.685 & 42.833 & 3.267 & 3.811 & 3.443 & 3.618 & 3.718 & 3.453 & 7.840 & 43.955 & 2.058 & -0.723 \\
37 & 020332.49-050217.0 & 2.057 & 45.457 & - & - & - & - & 43.155 & 43.326 & 43.550 & - & - & - & 3.730 & 3.943 & 3.634 & 8.343 & 44.478 & 2.072 & -0.704 \\
38 & 020332.82-050944.5 & 1.353 & 44.788 & - & - & - & - & 42.408 & 42.419 & 42.715 & - & - & - & 3.612 & 3.804 & 3.465 & 7.295 & 43.702 & 2.074 & -0.432 \\
39 & 020334.58-051721.3 & 1.131 & 43.425 & - & - & - & - & 41.338 & 41.548 & 41.757 & - & - & - & 3.303 & 3.804 & 3.720 & 7.350 & 44.216 & 2.017 & 0.027 \\
40 & 020334.79-063832.3 & 0.809 & 44.173 & 43.960 & 41.443 & 42.014 & 42.014 & 41.861 & 42.349 & 42.472 & 3.162 & 3.764 & 3.764 & 3.699 & 3.821 & 3.614 & 8.565 & 43.356 & 2.019 & -2.048 \\
41 & 020336.11-050106.7 & 1.228 & 44.786 & - & - & - & - & 42.534 & 42.695 & 42.923 & - & - & - & 3.705 & 3.938 & 3.644 & 8.315 & 44.066 & 2.114 & -1.088 \\
42 & 020336.45-051545.4 & 1.401 & 44.904 & - & - & - & - & 42.625 & 42.833 & 43.042 & - & - & - & 3.715 & 3.806 & 3.492 & 7.763 & 44.032 & 2.069 & -0.570 \\
43 & 020337.29-051406.4 & 0.519 & 44.204 & 44.030 & 41.771 & 42.183 & 42.325 & 42.137 & 42.339 & 42.551 & 3.183 & 3.580 & 3.359 & 3.374 & 3.679 & 3.362 & 6.886 & 43.804 & 1.906 & 0.079 \\
44 & 020337.36-042823.5 & 1.979 & 44.764 & - & - & - & - & 42.767 & 43.412 & 43.501 & - & - & - & 3.690 & 3.806 & 3.694 & 8.047 & 44.184 & 2.026 & -0.702 \\
45 & 020338.11-043903.2 & 1.770 & 44.783 & - & - & - & - & 42.427 & 43.031 & 43.128 & - & - & - & 3.607 & 3.824 & 3.666 & 8.696 & 44.132 & 1.973 & -1.403 \\

\hline
\end{tabular}
\end{table}
\end{landscape}










\bibliographystyle{harv}

\bibliography{cas-refs}

\begin{thebibliography}{75}
\expandafter\ifx\csname natexlab\endcsname\relax\def\natexlab#1{#1}\fi
\providecommand{\url}[1]{\texttt{#1}}
\providecommand{\href}[2]{#2}
\providecommand{\path}[1]{#1}
\providecommand{\DOIprefix}{doi:}
\providecommand{\ArXivprefix}{arXiv:}
\providecommand{\URLprefix}{URL: }
\providecommand{\Pubmedprefix}{pmid:}
\providecommand{\doi}[1]{\href{http://dx.doi.org/#1}{\path{#1}}}
\providecommand{\Pubmed}[1]{\href{pmid:#1}{\path{#1}}}
\providecommand{\bibinfo}[2]{#2}
\ifx\xfnm\relax \def\xfnm[#1]{\unskip,\space#1}\fi
\bibitem[{{Abolmasov} and {Poutanen}(2017)}]{Abolmasov+etal+2017}
\bibinfo{author}{{Abolmasov}, P.}, \bibinfo{author}{{Poutanen}, J.},
  \bibinfo{year}{2017}.
\newblock \bibinfo{title}{{Gamma-ray opacity of the anisotropic stratified
  broad-line regions in blazars}}.
\newblock \bibinfo{journal}{\mnras} \bibinfo{volume}{464},
  \bibinfo{pages}{152--169}.
\newblock \DOIprefix\doi{10.1093/mnras/stw2326},
  \href{http://arxiv.org/abs/1609.03350}{{\tt arXiv:1609.03350}}.
\bibitem[{{Antonucci}(1993)}]{Antonucci+1993}
\bibinfo{author}{{Antonucci}, R.}, \bibinfo{year}{1993}.
\newblock \bibinfo{title}{{Unified models for active galactic nuclei and
  quasars.}}
\newblock \bibinfo{journal}{\araa} \bibinfo{volume}{31},
  \bibinfo{pages}{473--521}.
\newblock \DOIprefix\doi{10.1146/annurev.aa.31.090193.002353}.
\bibitem[{{Antonucci}(1984)}]{Antonucci+1984}
\bibinfo{author}{{Antonucci}, R.R.J.}, \bibinfo{year}{1984}.
\newblock \bibinfo{title}{{Optical spectropolarimetry of radio galaxies.}}
\newblock \bibinfo{journal}{\apj} \bibinfo{volume}{278},
  \bibinfo{pages}{499--520}.
\newblock \DOIprefix\doi{10.1086/161816}.
\bibitem[{{Antonucci} and {Miller}(1985)}]{Antonucci+etal+1985}
\bibinfo{author}{{Antonucci}, R.R.J.}, \bibinfo{author}{{Miller}, J.S.},
  \bibinfo{year}{1985}.
\newblock \bibinfo{title}{{Spectropolarimetry and the nature of NGC 1068.}}
\newblock \bibinfo{journal}{\apj} \bibinfo{volume}{297},
  \bibinfo{pages}{621--632}.
\newblock \DOIprefix\doi{10.1086/163559}.
\bibitem[{{Assef} et~al.(2011){Assef}, {Denney}, {Kochanek}, {Peterson},
  {Koz{\l}owski}, {Ageorges}, {Barrows}, {Buschkamp}, {Dietrich}, {Falco},
  {Feiz}, {Gemperlein}, {Germeroth}, {Grier}, {Hofmann}, {Juette}, {Khan},
  {Kilic}, {Knierim}, {Laun}, {Lederer}, {Lehmitz}, {Lenzen}, {Mall}, {Madsen},
  {Mandel}, {Martini}, {Mathur}, {Mogren}, {Mueller}, {Naranjo}, {Pasquali},
  {Polsterer}, {Pogge}, {Quirrenbach}, {Seifert}, {Stern}, {Shappee}, {Storz},
  {Van Saders}, {Weiser} and {Zhang}}]{Assef+etal+2011}
\bibinfo{author}{{Assef}, R.J.}, \bibinfo{author}{{Denney}, K.D.},
  \bibinfo{author}{{Kochanek}, C.S.}, \bibinfo{author}{{Peterson}, B.M.},
  \bibinfo{author}{{Koz{\l}owski}, S.}, \bibinfo{author}{{Ageorges}, N.},
  \bibinfo{author}{{Barrows}, R.S.}, \bibinfo{author}{{Buschkamp}, P.},
  \bibinfo{author}{{Dietrich}, M.}, \bibinfo{author}{{Falco}, E.},
  \bibinfo{author}{{Feiz}, C.}, \bibinfo{author}{{Gemperlein}, H.},
  \bibinfo{author}{{Germeroth}, A.}, \bibinfo{author}{{Grier}, C.J.},
  \bibinfo{author}{{Hofmann}, R.}, \bibinfo{author}{{Juette}, M.},
  \bibinfo{author}{{Khan}, R.}, \bibinfo{author}{{Kilic}, M.},
  \bibinfo{author}{{Knierim}, V.}, \bibinfo{author}{{Laun}, W.},
  \bibinfo{author}{{Lederer}, R.}, \bibinfo{author}{{Lehmitz}, M.},
  \bibinfo{author}{{Lenzen}, R.}, \bibinfo{author}{{Mall}, U.},
  \bibinfo{author}{{Madsen}, K.K.}, \bibinfo{author}{{Mandel}, H.},
  \bibinfo{author}{{Martini}, P.}, \bibinfo{author}{{Mathur}, S.},
  \bibinfo{author}{{Mogren}, K.}, \bibinfo{author}{{Mueller}, P.},
  \bibinfo{author}{{Naranjo}, V.}, \bibinfo{author}{{Pasquali}, A.},
  \bibinfo{author}{{Polsterer}, K.}, \bibinfo{author}{{Pogge}, R.W.},
  \bibinfo{author}{{Quirrenbach}, A.}, \bibinfo{author}{{Seifert}, W.},
  \bibinfo{author}{{Stern}, D.}, \bibinfo{author}{{Shappee}, B.},
  \bibinfo{author}{{Storz}, C.}, \bibinfo{author}{{Van Saders}, J.},
  \bibinfo{author}{{Weiser}, P.}, \bibinfo{author}{{Zhang}, D.},
  \bibinfo{year}{2011}.
\newblock \bibinfo{title}{{Black Hole Mass Estimates Based on C IV are
  Consistent with Those Based on the Balmer Lines}}.
\newblock \bibinfo{journal}{\apj} \bibinfo{volume}{742}, \bibinfo{pages}{93}.
\newblock \DOIprefix\doi{10.1088/0004-637X/742/2/93},
  \href{http://arxiv.org/abs/1009.1145}{{\tt arXiv:1009.1145}}.
\bibitem[{{Barret} et~al.(2018){Barret}, {Lam Trong}, {den Herder}, {Piro},
  {Cappi}, {Houvelin}, {Kelley}, {Mas-Hesse}, {Mitsuda}, {Paltani}, {Rauw},
  {Rozanska}, {Wilms}, {Bandler}, {Barbera}, {Barcons}, {Bozzo}, {Ceballos},
  {Charles}, {Costantini}, {Decourchelle}, {den Hartog}, {Duband}, {Duval},
  {Fiore}, {Gatti}, {Goldwurm}, {Jackson}, {Jonker}, {Kilbourne}, {Macculi},
  {Mendez}, {Molendi}, {Orleanski}, {Pajot}, {Pointecouteau}, {Porter},
  {Pratt}, {Pr{\^e}le}, {Ravera}, {Sato}, {Schaye}, {Shinozaki}, {Thibert},
  {Valenziano}, {Valette}, {Vink}, {Webb}, {Wise}, {Yamasaki}, {Douchin},
  {Mesnager}, {Pontet}, {Pradines}, {Branduardi-Raymont}, {Bulbul}, {Dadina},
  {Ettori}, {Finoguenov}, {Fukazawa}, {Janiuk}, {Kaastra}, {Mazzotta},
  {Miller}, {Miniutti}, {Naze}, {Nicastro}, {Scioritino}, {Simonescu},
  {Torrejon}, {Frezouls}, {Geoffray}, {Peille}, {Aicardi}, {Andr{\'e}},
  {Daniel}, {Cl{\'e}net}, {Etcheverry}, {Gloaguen}, {Hervet}, {Jolly}, {Ledot},
  {Paillet}, {Schmisser}, {Vella}, {Damery}, {Boyce}, {Dipirro}, {Lotti},
  {Schwander}, {Smith}, {Van Leeuwen}, {van Weers}, {Clerc}, {Cobo}, {Dauser},
  {Kirsch}, {Cucchetti}, {Eckart}, {Ferrando} and
  {Natalucci}}]{Barret+etal+2018}
\bibinfo{author}{{Barret}, D.}, \bibinfo{author}{{Lam Trong}, T.},
  \bibinfo{author}{{den Herder}, J.W.}, \bibinfo{author}{{Piro}, L.},
  \bibinfo{author}{{Cappi}, M.}, \bibinfo{author}{{Houvelin}, J.},
  \bibinfo{author}{{Kelley}, R.}, \bibinfo{author}{{Mas-Hesse}, J.M.},
  \bibinfo{author}{{Mitsuda}, K.}, \bibinfo{author}{{Paltani}, S.},
  \bibinfo{author}{{Rauw}, G.}, \bibinfo{author}{{Rozanska}, A.},
  \bibinfo{author}{{Wilms}, J.}, \bibinfo{author}{{Bandler}, S.},
  \bibinfo{author}{{Barbera}, M.}, \bibinfo{author}{{Barcons}, X.},
  \bibinfo{author}{{Bozzo}, E.}, \bibinfo{author}{{Ceballos}, M.T.},
  \bibinfo{author}{{Charles}, I.}, \bibinfo{author}{{Costantini}, E.},
  \bibinfo{author}{{Decourchelle}, A.}, \bibinfo{author}{{den Hartog}, R.},
  \bibinfo{author}{{Duband}, L.}, \bibinfo{author}{{Duval}, J.M.},
  \bibinfo{author}{{Fiore}, F.}, \bibinfo{author}{{Gatti}, F.},
  \bibinfo{author}{{Goldwurm}, A.}, \bibinfo{author}{{Jackson}, B.},
  \bibinfo{author}{{Jonker}, P.}, \bibinfo{author}{{Kilbourne}, C.},
  \bibinfo{author}{{Macculi}, C.}, \bibinfo{author}{{Mendez}, M.},
  \bibinfo{author}{{Molendi}, S.}, \bibinfo{author}{{Orleanski}, P.},
  \bibinfo{author}{{Pajot}, F.}, \bibinfo{author}{{Pointecouteau}, E.},
  \bibinfo{author}{{Porter}, F.}, \bibinfo{author}{{Pratt}, G.W.},
  \bibinfo{author}{{Pr{\^e}le}, D.}, \bibinfo{author}{{Ravera}, L.},
  \bibinfo{author}{{Sato}, K.}, \bibinfo{author}{{Schaye}, J.},
  \bibinfo{author}{{Shinozaki}, K.}, \bibinfo{author}{{Thibert}, T.},
  \bibinfo{author}{{Valenziano}, L.}, \bibinfo{author}{{Valette}, V.},
  \bibinfo{author}{{Vink}, J.}, \bibinfo{author}{{Webb}, N.},
  \bibinfo{author}{{Wise}, M.}, \bibinfo{author}{{Yamasaki}, N.},
  \bibinfo{author}{{Douchin}, F.}, \bibinfo{author}{{Mesnager}, J.M.},
  \bibinfo{author}{{Pontet}, B.}, \bibinfo{author}{{Pradines}, A.},
  \bibinfo{author}{{Branduardi-Raymont}, G.}, \bibinfo{author}{{Bulbul}, E.},
  \bibinfo{author}{{Dadina}, M.}, \bibinfo{author}{{Ettori}, S.},
  \bibinfo{author}{{Finoguenov}, A.}, \bibinfo{author}{{Fukazawa}, Y.},
  \bibinfo{author}{{Janiuk}, A.}, \bibinfo{author}{{Kaastra}, J.},
  \bibinfo{author}{{Mazzotta}, P.}, \bibinfo{author}{{Miller}, J.},
  \bibinfo{author}{{Miniutti}, G.}, \bibinfo{author}{{Naze}, Y.},
  \bibinfo{author}{{Nicastro}, F.}, \bibinfo{author}{{Scioritino}, S.},
  \bibinfo{author}{{Simonescu}, A.}, \bibinfo{author}{{Torrejon}, J.M.},
  \bibinfo{author}{{Frezouls}, B.}, \bibinfo{author}{{Geoffray}, H.},
  \bibinfo{author}{{Peille}, P.}, \bibinfo{author}{{Aicardi}, C.},
  \bibinfo{author}{{Andr{\'e}}, J.}, \bibinfo{author}{{Daniel}, C.},
  \bibinfo{author}{{Cl{\'e}net}, A.}, \bibinfo{author}{{Etcheverry}, C.},
  \bibinfo{author}{{Gloaguen}, E.}, \bibinfo{author}{{Hervet}, G.},
  \bibinfo{author}{{Jolly}, A.}, \bibinfo{author}{{Ledot}, A.},
  \bibinfo{author}{{Paillet}, I.}, \bibinfo{author}{{Schmisser}, R.},
  \bibinfo{author}{{Vella}, B.}, \bibinfo{author}{{Damery}, J.C.},
  \bibinfo{author}{{Boyce}, K.}, \bibinfo{author}{{Dipirro}, M.},
  \bibinfo{author}{{Lotti}, S.}, \bibinfo{author}{{Schwander}, D.},
  \bibinfo{author}{{Smith}, S.}, \bibinfo{author}{{Van Leeuwen}, B.J.},
  \bibinfo{author}{{van Weers}, H.}, \bibinfo{author}{{Clerc}, N.},
  \bibinfo{author}{{Cobo}, B.}, \bibinfo{author}{{Dauser}, T.},
  \bibinfo{author}{{Kirsch}, C.}, \bibinfo{author}{{Cucchetti}, E.},
  \bibinfo{author}{{Eckart}, M.}, \bibinfo{author}{{Ferrando}, P.},
  \bibinfo{author}{{Natalucci}, L.}, \bibinfo{year}{2018}.
\newblock \bibinfo{title}{{The ATHENA X-ray Integral Field Unit (X-IFU)}}, in:
  \bibinfo{editor}{{den Herder}, J.W.A.}, \bibinfo{editor}{{Nikzad}, S.},
  \bibinfo{editor}{{Nakazawa}, K.} (Eds.), \bibinfo{booktitle}{Space Telescopes
  and Instrumentation 2018: Ultraviolet to Gamma Ray}, p.
  \bibinfo{pages}{106991G}.
\newblock \DOIprefix\doi{10.1117/12.2312409},
  \href{http://arxiv.org/abs/1807.06092}{{\tt arXiv:1807.06092}}.
\bibitem[{{Bianchi} et~al.(2009a){Bianchi}, {Bonilla}, {Guainazzi}, {Matt} and
  {Ponti}}]{Bianchi+etal+2009a}
\bibinfo{author}{{Bianchi}, S.}, \bibinfo{author}{{Bonilla}, N.F.},
  \bibinfo{author}{{Guainazzi}, M.}, \bibinfo{author}{{Matt}, G.},
  \bibinfo{author}{{Ponti}, G.}, \bibinfo{year}{2009}a.
\newblock \bibinfo{title}{{CAIXA: a catalogue of AGN in the XMM-Newton archive.
  II. Multiwavelength correlations}}.
\newblock \bibinfo{journal}{\aap} \bibinfo{volume}{501},
  \bibinfo{pages}{915--924}.
\newblock \DOIprefix\doi{10.1051/0004-6361/200911905},
  \href{http://arxiv.org/abs/0905.0267}{{\tt arXiv:0905.0267}}.
\bibitem[{{Bianchi} et~al.(2009b){Bianchi}, {Guainazzi}, {Matt}, {Fonseca
  Bonilla} and {Ponti}}]{Bianchi+etal+2009b}
\bibinfo{author}{{Bianchi}, S.}, \bibinfo{author}{{Guainazzi}, M.},
  \bibinfo{author}{{Matt}, G.}, \bibinfo{author}{{Fonseca Bonilla}, N.},
  \bibinfo{author}{{Ponti}, G.}, \bibinfo{year}{2009}b.
\newblock \bibinfo{title}{{CAIXA: a catalogue of AGN in the XMM-Newton archive.
  I. Spectral analysis}}.
\newblock \bibinfo{journal}{\aap} \bibinfo{volume}{495},
  \bibinfo{pages}{421--430}.
\newblock \DOIprefix\doi{10.1051/0004-6361:200810620},
  \href{http://arxiv.org/abs/0811.1126}{{\tt arXiv:0811.1126}}.
\bibitem[{{Blandford} et~al.(1990){Blandford}, Netzer, Woltjer, Courvoisier and
  Mayor}]{Blandford+etal+1990}
\bibinfo{author}{{Blandford}, R.D.}, \bibinfo{author}{Netzer, H.},
  \bibinfo{author}{Woltjer, L.}, \bibinfo{author}{Courvoisier, T.J.L.},
  \bibinfo{author}{Mayor, M.}, \bibinfo{year}{1990}.
\newblock \bibinfo{title}{{Active Galactic Nuclei}} .
\bibitem[{{Bon} et~al.(2009){Bon}, {Popovi{\'c}}, {Gavrilovi{\'c}}, {La Mura}
  and {Mediavilla}}]{Bon+etal+2009}
\bibinfo{author}{{Bon}, E.}, \bibinfo{author}{{Popovi{\'c}}, L.{\v{C}}.},
  \bibinfo{author}{{Gavrilovi{\'c}}, N.}, \bibinfo{author}{{La Mura}, G.},
  \bibinfo{author}{{Mediavilla}, E.}, \bibinfo{year}{2009}.
\newblock \bibinfo{title}{{Contribution of a disc component to single-peaked
  broad lines of active galactic nuclei}}.
\newblock \bibinfo{journal}{\mnras} \bibinfo{volume}{400},
  \bibinfo{pages}{924--936}.
\newblock \DOIprefix\doi{10.1111/j.1365-2966.2009.15511.x},
  \href{http://arxiv.org/abs/0908.2939}{{\tt arXiv:0908.2939}}.
\bibitem[{{Brightman} and {Nandra}(2011)}]{Brightman+etal+2011}
\bibinfo{author}{{Brightman}, M.}, \bibinfo{author}{{Nandra}, K.},
  \bibinfo{year}{2011}.
\newblock \bibinfo{title}{{An XMM-Newton spectral survey of 12
  {\ensuremath{\mu}}m selected galaxies - I. X-ray data}}.
\newblock \bibinfo{journal}{\mnras} \bibinfo{volume}{413},
  \bibinfo{pages}{1206--1235}.
\newblock \DOIprefix\doi{10.1111/j.1365-2966.2011.18207.x},
  \href{http://arxiv.org/abs/1012.3345}{{\tt arXiv:1012.3345}}.
\bibitem[{{Brightman} et~al.(2013){Brightman}, {Silverman}, {Mainieri}, {Ueda},
  {Schramm}, {Matsuoka}, {Nagao}, {Steinhardt}, {Kartaltepe}, {Sanders},
  {Treister}, {Shemmer}, {Brandt}, {Brusa}, {Comastri}, {Ho}, {Lanzuisi},
  {Lusso}, {Nandra}, {Salvato}, {Zamorani}, {Akiyama}, {Alexander},
  {Bongiorno}, {Capak}, {Civano}, {Del Moro}, {Doi}, {Elvis}, {Hasinger},
  {Laird}, {Masters}, {Mignoli}, {Ohta}, {Schawinski} and
  {Taniguchi}}]{Brightman+etal+2013}
\bibinfo{author}{{Brightman}, M.}, \bibinfo{author}{{Silverman}, J.D.},
  \bibinfo{author}{{Mainieri}, V.}, \bibinfo{author}{{Ueda}, Y.},
  \bibinfo{author}{{Schramm}, M.}, \bibinfo{author}{{Matsuoka}, K.},
  \bibinfo{author}{{Nagao}, T.}, \bibinfo{author}{{Steinhardt}, C.},
  \bibinfo{author}{{Kartaltepe}, J.}, \bibinfo{author}{{Sanders}, D.B.},
  \bibinfo{author}{{Treister}, E.}, \bibinfo{author}{{Shemmer}, O.},
  \bibinfo{author}{{Brandt}, W.N.}, \bibinfo{author}{{Brusa}, M.},
  \bibinfo{author}{{Comastri}, A.}, \bibinfo{author}{{Ho}, L.C.},
  \bibinfo{author}{{Lanzuisi}, G.}, \bibinfo{author}{{Lusso}, E.},
  \bibinfo{author}{{Nandra}, K.}, \bibinfo{author}{{Salvato}, M.},
  \bibinfo{author}{{Zamorani}, G.}, \bibinfo{author}{{Akiyama}, M.},
  \bibinfo{author}{{Alexander}, D.M.}, \bibinfo{author}{{Bongiorno}, A.},
  \bibinfo{author}{{Capak}, P.}, \bibinfo{author}{{Civano}, F.},
  \bibinfo{author}{{Del Moro}, A.}, \bibinfo{author}{{Doi}, A.},
  \bibinfo{author}{{Elvis}, M.}, \bibinfo{author}{{Hasinger}, G.},
  \bibinfo{author}{{Laird}, E.S.}, \bibinfo{author}{{Masters}, D.},
  \bibinfo{author}{{Mignoli}, M.}, \bibinfo{author}{{Ohta}, K.},
  \bibinfo{author}{{Schawinski}, K.}, \bibinfo{author}{{Taniguchi}, Y.},
  \bibinfo{year}{2013}.
\newblock \bibinfo{title}{{A statistical relation between the X-ray spectral
  index and Eddington ratio of active galactic nuclei in deep surveys}}.
\newblock \bibinfo{journal}{\mnras} \bibinfo{volume}{433},
  \bibinfo{pages}{2485--2496}.
\newblock \DOIprefix\doi{10.1093/mnras/stt920},
  \href{http://arxiv.org/abs/1305.3917}{{\tt arXiv:1305.3917}}.
\bibitem[{{Brinkmann} et~al.(1997){Brinkmann}, {Yuan} and
  {Siebert}}]{Brinkmann+etal+1997}
\bibinfo{author}{{Brinkmann}, W.}, \bibinfo{author}{{Yuan}, W.},
  \bibinfo{author}{{Siebert}, J.}, \bibinfo{year}{1997}.
\newblock \bibinfo{title}{{Broad band energy distribution of ROSAT detected
  quasars. I. Radio-loud objects.}}
\newblock \bibinfo{journal}{\aap} \bibinfo{volume}{319},
  \bibinfo{pages}{413--429}.
\bibitem[{{Buchner} et~al.(2014){Buchner}, {Georgakakis}, {Nandra}, {Hsu},
  {Rangel}, {Brightman}, {Merloni}, {Salvato}, {Donley} and
  {Kocevski}}]{Buchner+etal+2014}
\bibinfo{author}{{Buchner}, J.}, \bibinfo{author}{{Georgakakis}, A.},
  \bibinfo{author}{{Nandra}, K.}, \bibinfo{author}{{Hsu}, L.},
  \bibinfo{author}{{Rangel}, C.}, \bibinfo{author}{{Brightman}, M.},
  \bibinfo{author}{{Merloni}, A.}, \bibinfo{author}{{Salvato}, M.},
  \bibinfo{author}{{Donley}, J.}, \bibinfo{author}{{Kocevski}, D.},
  \bibinfo{year}{2014}.
\newblock \bibinfo{title}{{X-ray spectral modelling of the AGN obscuring region
  in the CDFS: Bayesian model selection and catalogue}}.
\newblock \bibinfo{journal}{\aap} \bibinfo{volume}{564}, \bibinfo{pages}{A125}.
\newblock \DOIprefix\doi{10.1051/0004-6361/201322971},
  \href{http://arxiv.org/abs/1402.0004}{{\tt arXiv:1402.0004}}.
\bibitem[{{Chartas} et~al.(2009){Chartas}, {Kochanek}, {Dai}, {Poindexter} and
  {Garmire}}]{Chartas+etal+2009}
\bibinfo{author}{{Chartas}, G.}, \bibinfo{author}{{Kochanek}, C.S.},
  \bibinfo{author}{{Dai}, X.}, \bibinfo{author}{{Poindexter}, S.},
  \bibinfo{author}{{Garmire}, G.}, \bibinfo{year}{2009}.
\newblock \bibinfo{title}{{X-Ray Microlensing in RXJ1131-1231 and
  HE1104-1805}}.
\newblock \bibinfo{journal}{\apj} \bibinfo{volume}{693},
  \bibinfo{pages}{174--185}.
\newblock \DOIprefix\doi{10.1088/0004-637X/693/1/174},
  \href{http://arxiv.org/abs/0805.4492}{{\tt arXiv:0805.4492}}.
\bibitem[{{De Marco} et~al.(2013){De Marco}, {Ponti}, {Cappi}, {Dadina},
  {Uttley}, {Cackett}, {Fabian} and {Miniutti}}]{DeMarco+etal+2013}
\bibinfo{author}{{De Marco}, B.}, \bibinfo{author}{{Ponti}, G.},
  \bibinfo{author}{{Cappi}, M.}, \bibinfo{author}{{Dadina}, M.},
  \bibinfo{author}{{Uttley}, P.}, \bibinfo{author}{{Cackett}, E.M.},
  \bibinfo{author}{{Fabian}, A.C.}, \bibinfo{author}{{Miniutti}, G.},
  \bibinfo{year}{2013}.
\newblock \bibinfo{title}{{Discovery of a relation between black hole mass and
  soft X-ray time lags in active galactic nuclei}}.
\newblock \bibinfo{journal}{\mnras} \bibinfo{volume}{431},
  \bibinfo{pages}{2441--2452}.
\newblock \DOIprefix\doi{10.1093/mnras/stt339},
  \href{http://arxiv.org/abs/1201.0196}{{\tt arXiv:1201.0196}}.
\bibitem[{{Fabian} et~al.(2009){Fabian}, {Zoghbi}, {Ross}, {Uttley}, {Gallo},
  {Brandt}, {Blustin}, {Boller}, {Caballero-Garcia}, {Larsson}, {Miller},
  {Miniutti}, {Ponti}, {Reis}, {Reynolds}, {Tanaka} and
  {Young}}]{Fabian+etal+2009}
\bibinfo{author}{{Fabian}, A.C.}, \bibinfo{author}{{Zoghbi}, A.},
  \bibinfo{author}{{Ross}, R.R.}, \bibinfo{author}{{Uttley}, P.},
  \bibinfo{author}{{Gallo}, L.C.}, \bibinfo{author}{{Brandt}, W.N.},
  \bibinfo{author}{{Blustin}, A.J.}, \bibinfo{author}{{Boller}, T.},
  \bibinfo{author}{{Caballero-Garcia}, M.D.}, \bibinfo{author}{{Larsson}, J.},
  \bibinfo{author}{{Miller}, J.M.}, \bibinfo{author}{{Miniutti}, G.},
  \bibinfo{author}{{Ponti}, G.}, \bibinfo{author}{{Reis}, R.C.},
  \bibinfo{author}{{Reynolds}, C.S.}, \bibinfo{author}{{Tanaka}, Y.},
  \bibinfo{author}{{Young}, A.J.}, \bibinfo{year}{2009}.
\newblock \bibinfo{title}{{Broad line emission from iron K- and L-shell
  transitions in the active galaxy 1H0707-495}}.
\newblock \bibinfo{journal}{\nat} \bibinfo{volume}{459},
  \bibinfo{pages}{540--542}.
\newblock \DOIprefix\doi{10.1038/nature08007}.
\bibitem[{{Fanali} et~al.(2013){Fanali}, {Caccianiga}, {Severgnini}, {Della
  Ceca}, {Marchese}, {Carrera}, {Corral} and {Mateos}}]{Fanali+etal+2013}
\bibinfo{author}{{Fanali}, R.}, \bibinfo{author}{{Caccianiga}, A.},
  \bibinfo{author}{{Severgnini}, P.}, \bibinfo{author}{{Della Ceca}, R.},
  \bibinfo{author}{{Marchese}, E.}, \bibinfo{author}{{Carrera}, F.J.},
  \bibinfo{author}{{Corral}, A.}, \bibinfo{author}{{Mateos}, S.},
  \bibinfo{year}{2013}.
\newblock \bibinfo{title}{{Studying the relationship between X-ray emission and
  accretion in AGN using the XMM-Newton Bright Serendipitous Survey}}.
\newblock \bibinfo{journal}{\mnras} \bibinfo{volume}{433},
  \bibinfo{pages}{648--658}.
\newblock \DOIprefix\doi{10.1093/mnras/stt757},
  \href{http://arxiv.org/abs/1305.0564}{{\tt arXiv:1305.0564}}.
\bibitem[{{Fitzpatrick}(1999)}]{Fitzpatrick+etal+1999}
\bibinfo{author}{{Fitzpatrick}, E.L.}, \bibinfo{year}{1999}.
\newblock \bibinfo{title}{{Correcting for the Effects of Interstellar
  Extinction}}.
\newblock \bibinfo{journal}{\pasp} \bibinfo{volume}{111},
  \bibinfo{pages}{63--75}.
\newblock \DOIprefix\doi{10.1086/316293},
  \href{http://arxiv.org/abs/astro-ph/9809387}{{\tt arXiv:astro-ph/9809387}}.
\bibitem[{{Freeman} et~al.(2001){Freeman}, {Doe} and
  {Siemiginowska}}]{Freeman+etal+2001}
\bibinfo{author}{{Freeman}, P.}, \bibinfo{author}{{Doe}, S.},
  \bibinfo{author}{{Siemiginowska}, A.}, \bibinfo{year}{2001}.
\newblock \bibinfo{title}{{Sherpa: a mission-independent data analysis
  application}}, in: \bibinfo{editor}{{Starck}, J.L.},
  \bibinfo{editor}{{Murtagh}, F.D.} (Eds.), \bibinfo{booktitle}{Astronomical
  Data Analysis}, pp. \bibinfo{pages}{76--87}.
\newblock \DOIprefix\doi{10.1117/12.447161},
  \href{http://arxiv.org/abs/astro-ph/0108426}{{\tt arXiv:astro-ph/0108426}}.
\bibitem[{{Haardt} and {Maraschi}(1991)}]{Haardt+etal+1991}
\bibinfo{author}{{Haardt}, F.}, \bibinfo{author}{{Maraschi}, L.},
  \bibinfo{year}{1991}.
\newblock \bibinfo{title}{{A Two-Phase Model for the X-Ray Emission from
  Seyfert Galaxies}}.
\newblock \bibinfo{journal}{\apjl} \bibinfo{volume}{380}, \bibinfo{pages}{L51}.
\newblock \DOIprefix\doi{10.1086/186171}.
\bibitem[{{Haardt} and {Maraschi}(1993)}]{Haardt+etal+1993}
\bibinfo{author}{{Haardt}, F.}, \bibinfo{author}{{Maraschi}, L.},
  \bibinfo{year}{1993}.
\newblock \bibinfo{title}{{X-Ray Spectra from Two-Phase Accretion Disks}}.
\newblock \bibinfo{journal}{\apj} \bibinfo{volume}{413}, \bibinfo{pages}{507}.
\newblock \DOIprefix\doi{10.1086/173020}.
\bibitem[{{Jin} et~al.(2012a){Jin}, {Ward} and {Done}}]{Jin+etal+2012a}
\bibinfo{author}{{Jin}, C.}, \bibinfo{author}{{Ward}, M.},
  \bibinfo{author}{{Done}, C.}, \bibinfo{year}{2012}a.
\newblock \bibinfo{title}{{A combined optical and X-ray study of unobscured
  type 1 active galactic nuclei - II. Relation between X-ray emission and
  optical spectra}}.
\newblock \bibinfo{journal}{\mnras} \bibinfo{volume}{422},
  \bibinfo{pages}{3268--3284}.
\newblock \DOIprefix\doi{10.1111/j.1365-2966.2012.20847.x},
  \href{http://arxiv.org/abs/1203.0239}{{\tt arXiv:1203.0239}}.
\bibitem[{{Jin} et~al.(2012b){Jin}, {Ward} and {Done}}]{Jin+atal+2012}
\bibinfo{author}{{Jin}, C.}, \bibinfo{author}{{Ward}, M.},
  \bibinfo{author}{{Done}, C.}, \bibinfo{year}{2012}b.
\newblock \bibinfo{title}{{A combined optical and X-ray study of unobscured
  type 1 active galactic nuclei - III. Broad-band SED properties}}.
\newblock \bibinfo{journal}{\mnras} \bibinfo{volume}{425},
  \bibinfo{pages}{907--929}.
\newblock \DOIprefix\doi{10.1111/j.1365-2966.2012.21272.x},
  \href{http://arxiv.org/abs/1205.1846}{{\tt arXiv:1205.1846}}.
\bibitem[{{Joni{\'c}} et~al.(2016){Joni{\'c}},
  {Kova{\v{c}}evi{\'c}-Doj{\v{c}}inovi{\'c}}, {Ili{\'c}} and
  {Popovi{\'c}}}]{Jonic+etal+2016}
\bibinfo{author}{{Joni{\'c}}, S.},
  \bibinfo{author}{{Kova{\v{c}}evi{\'c}-Doj{\v{c}}inovi{\'c}}, J.},
  \bibinfo{author}{{Ili{\'c}}, D.}, \bibinfo{author}{{Popovi{\'c}},
  L.{\v{C}}.}, \bibinfo{year}{2016}.
\newblock \bibinfo{title}{{Virilization of the Broad Line Region in Active
  Galactic Nuclei{\textemdash}connection between shifts and widths of broad
  emission lines}}.
\newblock \bibinfo{journal}{\apss} \bibinfo{volume}{361}, \bibinfo{pages}{101}.
\newblock \DOIprefix\doi{10.1007/s10509-016-2680-9},
  \href{http://arxiv.org/abs/1602.03668}{{\tt arXiv:1602.03668}}.
\bibitem[{{Kamraj} et~al.(2022){Kamraj}, {Brightman}, {Harrison}, {Stern},
  {Garc{\'\i}a}, {Balokovi{\'c}}, {Ricci}, {Koss}, {Mej{\'\i}a-Restrepo}, {Oh},
  {Powell} and {Urry}}]{Kamraj+etal+2022}
\bibinfo{author}{{Kamraj}, N.}, \bibinfo{author}{{Brightman}, M.},
  \bibinfo{author}{{Harrison}, F.A.}, \bibinfo{author}{{Stern}, D.},
  \bibinfo{author}{{Garc{\'\i}a}, J.A.}, \bibinfo{author}{{Balokovi{\'c}}, M.},
  \bibinfo{author}{{Ricci}, C.}, \bibinfo{author}{{Koss}, M.J.},
  \bibinfo{author}{{Mej{\'\i}a-Restrepo}, J.E.}, \bibinfo{author}{{Oh}, K.},
  \bibinfo{author}{{Powell}, M.C.}, \bibinfo{author}{{Urry}, C.M.},
  \bibinfo{year}{2022}.
\newblock \bibinfo{title}{{X-Ray Coronal Properties of Swift/BAT-selected
  Seyfert 1 Active Galactic Nuclei}}.
\newblock \bibinfo{journal}{\apj} \bibinfo{volume}{927}, \bibinfo{pages}{42}.
\newblock \DOIprefix\doi{10.3847/1538-4357/ac45f6},
  \href{http://arxiv.org/abs/2202.00895}{{\tt arXiv:2202.00895}}.
\bibitem[{{Kara} et~al.(2013){Kara}, {Fabian}, {Cackett}, {Uttley}, {Wilkins}
  and {Zoghbi}}]{Kara+etal+2013}
\bibinfo{author}{{Kara}, E.}, \bibinfo{author}{{Fabian}, A.C.},
  \bibinfo{author}{{Cackett}, E.M.}, \bibinfo{author}{{Uttley}, P.},
  \bibinfo{author}{{Wilkins}, D.R.}, \bibinfo{author}{{Zoghbi}, A.},
  \bibinfo{year}{2013}.
\newblock \bibinfo{title}{{Discovery of high-frequency iron K lags in Ark 564
  and Mrk 335}}.
\newblock \bibinfo{journal}{\mnras} \bibinfo{volume}{434},
  \bibinfo{pages}{1129--1137}.
\newblock \DOIprefix\doi{10.1093/mnras/stt1055},
  \href{http://arxiv.org/abs/1306.2551}{{\tt arXiv:1306.2551}}.
\bibitem[{{Kelly} et~al.(2007){Kelly}, {Bechtold}, {Siemiginowska}, {Aldcroft}
  and {Sobolewska}}]{Kelly+etal+2007}
\bibinfo{author}{{Kelly}, B.C.}, \bibinfo{author}{{Bechtold}, J.},
  \bibinfo{author}{{Siemiginowska}, A.}, \bibinfo{author}{{Aldcroft}, T.},
  \bibinfo{author}{{Sobolewska}, M.}, \bibinfo{year}{2007}.
\newblock \bibinfo{title}{{Evolution of the X-ray Emission of Radio-quiet
  Quasars}}.
\newblock \bibinfo{journal}{\apj} \bibinfo{volume}{657},
  \bibinfo{pages}{116--134}.
\newblock \DOIprefix\doi{10.1086/510876},
  \href{http://arxiv.org/abs/astro-ph/0611120}{{\tt arXiv:astro-ph/0611120}}.
\bibitem[{{Kova{\v{c}}evi{\'c}} et~al.(2010){Kova{\v{c}}evi{\'c}},
  {Popovi{\'c}} and {Dimitrijevi{\'c}}}]{Kova+etal+2010}
\bibinfo{author}{{Kova{\v{c}}evi{\'c}}, J.}, \bibinfo{author}{{Popovi{\'c}},
  L.{\v{C}}.}, \bibinfo{author}{{Dimitrijevi{\'c}}, M.S.},
  \bibinfo{year}{2010}.
\newblock \bibinfo{title}{{Analysis of Optical Fe II Emission in a Sample of
  Active Galactic Nucleus Spectra}}.
\newblock \bibinfo{journal}{\apjs} \bibinfo{volume}{189},
  \bibinfo{pages}{15--36}.
\newblock \DOIprefix\doi{10.1088/0067-0049/189/1/15},
  \href{http://arxiv.org/abs/1004.2212}{{\tt arXiv:1004.2212}}.
\bibitem[{{Kova{\v{c}}evi{\'c}-Doj{\v{c}}inovi{\'c}} and
  {Popovi{\'c}}(2015)}]{kova+etal+2015}
\bibinfo{author}{{Kova{\v{c}}evi{\'c}-Doj{\v{c}}inovi{\'c}}, J.},
  \bibinfo{author}{{Popovi{\'c}}, L.{\v{C}}.}, \bibinfo{year}{2015}.
\newblock \bibinfo{title}{{The Connections Between the UV and Optical Fe ii
  Emission Lines in Type 1 AGNs}}.
\newblock \bibinfo{journal}{\apjs} \bibinfo{volume}{221}, \bibinfo{pages}{35}.
\newblock \DOIprefix\doi{10.1088/0067-0049/221/2/35},
  \href{http://arxiv.org/abs/1509.03679}{{\tt arXiv:1509.03679}}.
\bibitem[{{Kuraszkiewicz} et~al.(2002){Kuraszkiewicz}, {Green}, {Forster},
  {Aldcroft}, {Evans} and {Koratkar}}]{Kuraszkiewicz+etal+2002}
\bibinfo{author}{{Kuraszkiewicz}, J.K.}, \bibinfo{author}{{Green}, P.J.},
  \bibinfo{author}{{Forster}, K.}, \bibinfo{author}{{Aldcroft}, T.L.},
  \bibinfo{author}{{Evans}, I.N.}, \bibinfo{author}{{Koratkar}, A.},
  \bibinfo{year}{2002}.
\newblock \bibinfo{title}{{Emission Line Properties of Active Galactic Nuclei
  from a pre-COSTAR Faint Object Spectrograph Hubble Space Telescope Spectral
  Atlas}}.
\newblock \bibinfo{journal}{\apjs} \bibinfo{volume}{143},
  \bibinfo{pages}{257--276}.
\newblock \DOIprefix\doi{10.1086/342789}.
\bibitem[{{Liu} et~al.(2017){Liu}, {Taam}, {Qiao} and {Yuan}}]{Liu+etal+2017}
\bibinfo{author}{{Liu}, B.F.}, \bibinfo{author}{{Taam}, R.E.},
  \bibinfo{author}{{Qiao}, E.}, \bibinfo{author}{{Yuan}, W.},
  \bibinfo{year}{2017}.
\newblock \bibinfo{title}{{Centrally Concentrated X-Ray Radiation from an
  Extended Accreting Corona in Active Galactic Nuclei}}.
\newblock \bibinfo{journal}{\apj} \bibinfo{volume}{847}, \bibinfo{pages}{96}.
\newblock \DOIprefix\doi{10.3847/1538-4357/aa894c},
  \href{http://arxiv.org/abs/1709.09799}{{\tt arXiv:1709.09799}}.
\bibitem[{{Liu} et~al.(2016){Liu}, {Merloni}, {Georgakakis}, {Menzel},
  {Buchner}, {Nandra}, {Salvato}, {Shen}, {Brusa} and
  {Streblyanska}}]{Liu+etal+2016}
\bibinfo{author}{{Liu}, Z.}, \bibinfo{author}{{Merloni}, A.},
  \bibinfo{author}{{Georgakakis}, A.}, \bibinfo{author}{{Menzel}, M.L.},
  \bibinfo{author}{{Buchner}, J.}, \bibinfo{author}{{Nandra}, K.},
  \bibinfo{author}{{Salvato}, M.}, \bibinfo{author}{{Shen}, Y.},
  \bibinfo{author}{{Brusa}, M.}, \bibinfo{author}{{Streblyanska}, A.},
  \bibinfo{year}{2016}.
\newblock \bibinfo{title}{{X-ray spectral properties of the AGN sample in the
  northern XMM-XXL field}}.
\newblock \bibinfo{journal}{\mnras} \bibinfo{volume}{459},
  \bibinfo{pages}{1602--1625}.
\newblock \DOIprefix\doi{10.1093/mnras/stw753},
  \href{http://arxiv.org/abs/1605.00207}{{\tt arXiv:1605.00207}}.
\bibitem[{{Marchese} et~al.(2012){Marchese}, {Della Ceca}, {Caccianiga},
  {Severgnini}, {Corral} and {Fanali}}]{Marchese+etal+2012}
\bibinfo{author}{{Marchese}, E.}, \bibinfo{author}{{Della Ceca}, R.},
  \bibinfo{author}{{Caccianiga}, A.}, \bibinfo{author}{{Severgnini}, P.},
  \bibinfo{author}{{Corral}, A.}, \bibinfo{author}{{Fanali}, R.},
  \bibinfo{year}{2012}.
\newblock \bibinfo{title}{{The optical-UV spectral energy distribution of the
  unabsorbed AGN population in the XMM-Newton Bright Serendipitous Survey}}.
\newblock \bibinfo{journal}{\aap} \bibinfo{volume}{539}, \bibinfo{pages}{A48}.
\newblock \DOIprefix\doi{10.1051/0004-6361/201117562},
  \href{http://arxiv.org/abs/1111.4409}{{\tt arXiv:1111.4409}}.
\bibitem[{{McHardy} et~al.(2005){McHardy}, {Gunn}, {Uttley} and
  {Goad}}]{McHardy+etal+2005}
\bibinfo{author}{{McHardy}, I.M.}, \bibinfo{author}{{Gunn}, K.F.},
  \bibinfo{author}{{Uttley}, P.}, \bibinfo{author}{{Goad}, M.R.},
  \bibinfo{year}{2005}.
\newblock \bibinfo{title}{{MCG-6-30-15: long time-scale X-ray variability,
  black hole mass and active galactic nuclei high states}}.
\newblock \bibinfo{journal}{\mnras} \bibinfo{volume}{359},
  \bibinfo{pages}{1469--1480}.
\newblock \DOIprefix\doi{10.1111/j.1365-2966.2005.08992.x},
  \href{http://arxiv.org/abs/astro-ph/0503100}{{\tt arXiv:astro-ph/0503100}}.
\bibitem[{{Nandra} et~al.(2007){Nandra}, {O'Neill}, {George} and
  {Reeves}}]{Nandra+etal+2007}
\bibinfo{author}{{Nandra}, K.}, \bibinfo{author}{{O'Neill}, P.M.},
  \bibinfo{author}{{George}, I.M.}, \bibinfo{author}{{Reeves}, J.N.},
  \bibinfo{year}{2007}.
\newblock \bibinfo{title}{{An XMM-Newton survey of broad iron lines in Seyfert
  galaxies}}.
\newblock \bibinfo{journal}{\mnras} \bibinfo{volume}{382},
  \bibinfo{pages}{194--228}.
\newblock \DOIprefix\doi{10.1111/j.1365-2966.2007.12331.x},
  \href{http://arxiv.org/abs/0708.1305}{{\tt arXiv:0708.1305}}.
\bibitem[{{Ojha} et~al.(2020){Ojha}, {Chand}, {Dewangan} and
  {Rakshit}}]{Ojha+etal+2020}
\bibinfo{author}{{Ojha}, V.}, \bibinfo{author}{{Chand}, H.},
  \bibinfo{author}{{Dewangan}, G.C.}, \bibinfo{author}{{Rakshit}, S.},
  \bibinfo{year}{2020}.
\newblock \bibinfo{title}{{A Comparison of X-Ray Photon Indices among the
  Narrow- and Broad-line Seyfert 1 Galaxies}}.
\newblock \bibinfo{journal}{\apj} \bibinfo{volume}{896}, \bibinfo{pages}{95}.
\newblock \DOIprefix\doi{10.3847/1538-4357/ab94ac},
  \href{http://arxiv.org/abs/2005.08352}{{\tt arXiv:2005.08352}}.
\bibitem[{{Page} et~al.(2004){Page}, {Reeves}, {O'Brien}, {Turner} and
  {Worrall}}]{Page+etal+2004}
\bibinfo{author}{{Page}, K.L.}, \bibinfo{author}{{Reeves}, J.N.},
  \bibinfo{author}{{O'Brien}, P.T.}, \bibinfo{author}{{Turner}, M.J.L.},
  \bibinfo{author}{{Worrall}, D.M.}, \bibinfo{year}{2004}.
\newblock \bibinfo{title}{{XMM-Newton observations of high-luminosity
  radio-quiet quasi-stellar objects}}.
\newblock \bibinfo{journal}{\mnras} \bibinfo{volume}{353},
  \bibinfo{pages}{133--142}.
\newblock \DOIprefix\doi{10.1111/j.1365-2966.2004.08051.x},
  \href{http://arxiv.org/abs/astro-ph/0405458}{{\tt arXiv:astro-ph/0405458}}.
\bibitem[{{P{\^a}ris} et~al.(2018){P{\^a}ris}, {Petitjean}, {Aubourg}, {Myers},
  {Streblyanska}, {Lyke}, {Anderson}, {Armengaud}, {Bautista}, {Blanton},
  {Blomqvist}, {Brinkmann}, {Brownstein}, {Brandt}, {Burtin}, {Dawson}, {de la
  Torre}, {Georgakakis}, {Gil-Mar{\'\i}n}, {Green}, {Hall}, {Kneib}, {LaMassa},
  {Le Goff}, {MacLeod}, {Mariappan}, {McGreer}, {Merloni}, {Noterdaeme},
  {Palanque-Delabrouille}, {Percival}, {Ross}, {Rossi}, {Schneider}, {Seo},
  {Tojeiro}, {Weaver}, {Weijmans}, {Y{\`e}che}, {Zarrouk} and
  {Zhao}}]{Paris+etal+2018}
\bibinfo{author}{{P{\^a}ris}, I.}, \bibinfo{author}{{Petitjean}, P.},
  \bibinfo{author}{{Aubourg}, {\'E}.}, \bibinfo{author}{{Myers}, A.D.},
  \bibinfo{author}{{Streblyanska}, A.}, \bibinfo{author}{{Lyke}, B.W.},
  \bibinfo{author}{{Anderson}, S.F.}, \bibinfo{author}{{Armengaud}, {\'E}.},
  \bibinfo{author}{{Bautista}, J.}, \bibinfo{author}{{Blanton}, M.R.},
  \bibinfo{author}{{Blomqvist}, M.}, \bibinfo{author}{{Brinkmann}, J.},
  \bibinfo{author}{{Brownstein}, J.R.}, \bibinfo{author}{{Brandt}, W.N.},
  \bibinfo{author}{{Burtin}, {\'E}.}, \bibinfo{author}{{Dawson}, K.},
  \bibinfo{author}{{de la Torre}, S.}, \bibinfo{author}{{Georgakakis}, A.},
  \bibinfo{author}{{Gil-Mar{\'\i}n}, H.}, \bibinfo{author}{{Green}, P.J.},
  \bibinfo{author}{{Hall}, P.B.}, \bibinfo{author}{{Kneib}, J.P.},
  \bibinfo{author}{{LaMassa}, S.M.}, \bibinfo{author}{{Le Goff}, J.M.},
  \bibinfo{author}{{MacLeod}, C.}, \bibinfo{author}{{Mariappan}, V.},
  \bibinfo{author}{{McGreer}, I.D.}, \bibinfo{author}{{Merloni}, A.},
  \bibinfo{author}{{Noterdaeme}, P.}, \bibinfo{author}{{Palanque-Delabrouille},
  N.}, \bibinfo{author}{{Percival}, W.J.}, \bibinfo{author}{{Ross}, A.J.},
  \bibinfo{author}{{Rossi}, G.}, \bibinfo{author}{{Schneider}, D.P.},
  \bibinfo{author}{{Seo}, H.J.}, \bibinfo{author}{{Tojeiro}, R.},
  \bibinfo{author}{{Weaver}, B.A.}, \bibinfo{author}{{Weijmans}, A.M.},
  \bibinfo{author}{{Y{\`e}che}, C.}, \bibinfo{author}{{Zarrouk}, P.},
  \bibinfo{author}{{Zhao}, G.B.}, \bibinfo{year}{2018}.
\newblock \bibinfo{title}{{The Sloan Digital Sky Survey Quasar Catalog:
  Fourteenth data release}}.
\newblock \bibinfo{journal}{\aap} \bibinfo{volume}{613}, \bibinfo{pages}{A51}.
\newblock \DOIprefix\doi{10.1051/0004-6361/201732445},
  \href{http://arxiv.org/abs/1712.05029}{{\tt arXiv:1712.05029}}.
\bibitem[{{Peterson}(1997)}]{Peterson+1997}
\bibinfo{author}{{Peterson}, B.M.}, \bibinfo{year}{1997}.
\newblock \bibinfo{title}{{An Introduction to Active Galactic Nuclei}}.
\bibitem[{{Piconcelli} et~al.(2005){Piconcelli}, {Jimenez-Bail{\'o}n},
  {Guainazzi}, {Schartel}, {Rodr{\'\i}guez-Pascual} and
  {Santos-Lle{\'o}}}]{Piconcelli+etal+2005}
\bibinfo{author}{{Piconcelli}, E.}, \bibinfo{author}{{Jimenez-Bail{\'o}n}, E.},
  \bibinfo{author}{{Guainazzi}, M.}, \bibinfo{author}{{Schartel}, N.},
  \bibinfo{author}{{Rodr{\'\i}guez-Pascual}, P.M.},
  \bibinfo{author}{{Santos-Lle{\'o}}, M.}, \bibinfo{year}{2005}.
\newblock \bibinfo{title}{{The XMM-Newton view of PG quasars. I. X-ray
  continuum and absorption}}.
\newblock \bibinfo{journal}{\aap} \bibinfo{volume}{432},
  \bibinfo{pages}{15--30}.
\newblock \DOIprefix\doi{10.1051/0004-6361:20041621},
  \href{http://arxiv.org/abs/astro-ph/0411051}{{\tt arXiv:astro-ph/0411051}}.
\bibitem[{Pierre et~al.(2016)Pierre, Pacaud, Adami, Alis, Altieri, Baran,
  Benoist, Birkinshaw, Bongiorno, Bremer et~al.}]{Pierre+etal+2016}
\bibinfo{author}{Pierre, M.}, \bibinfo{author}{Pacaud, F.},
  \bibinfo{author}{Adami, C.}, \bibinfo{author}{Alis, S.},
  \bibinfo{author}{Altieri, B.}, \bibinfo{author}{Baran, N.},
  \bibinfo{author}{Benoist, C.}, \bibinfo{author}{Birkinshaw, M.},
  \bibinfo{author}{Bongiorno, A.}, \bibinfo{author}{Bremer, M.}, et~al.,
  \bibinfo{year}{2016}.
\newblock \bibinfo{title}{The xxl survey-i. scientific motivations- xmm-newton
  observing plan- follow-up observations and simulation programme}.
\newblock \bibinfo{journal}{Astronomy \& Astrophysics} \bibinfo{volume}{592},
  \bibinfo{pages}{A1}.
\bibitem[{{Popovi{\'c}} et~al.(2019){Popovi{\'c}},
  {Kova{\v{c}}evi{\'c}-Doj{\v{c}}inovi{\'c}} and
  {Mar{\v{c}}eta-Mandi{\'c}}}]{Popovi+etal+2019}
\bibinfo{author}{{Popovi{\'c}}, L.{\v{C}}.},
  \bibinfo{author}{{Kova{\v{c}}evi{\'c}-Doj{\v{c}}inovi{\'c}}, J.},
  \bibinfo{author}{{Mar{\v{c}}eta-Mandi{\'c}}, S.}, \bibinfo{year}{2019}.
\newblock \bibinfo{title}{{The structure of the Mg II broad line emitting
  region in Type 1 AGNs}}.
\newblock \bibinfo{journal}{\mnras} \bibinfo{volume}{484},
  \bibinfo{pages}{3180--3197}.
\newblock \DOIprefix\doi{10.1093/mnras/stz157},
  \href{http://arxiv.org/abs/1901.03681}{{\tt arXiv:1901.03681}}.
\bibitem[{{Popovi{\'c}} et~al.(2004){Popovi{\'c}}, {Mediavilla}, {Bon} and
  {Ili{\'c}}}]{Popovic+etal+2004}
\bibinfo{author}{{Popovi{\'c}}, L.{\v{C}}.}, \bibinfo{author}{{Mediavilla},
  E.}, \bibinfo{author}{{Bon}, E.}, \bibinfo{author}{{Ili{\'c}}, D.},
  \bibinfo{year}{2004}.
\newblock \bibinfo{title}{{Contribution of the disk emission to the broad
  emission lines in AGNs: Two-component model}}.
\newblock \bibinfo{journal}{\aap} \bibinfo{volume}{423},
  \bibinfo{pages}{909--918}.
\newblock \DOIprefix\doi{10.1051/0004-6361:20034431},
  \href{http://arxiv.org/abs/astro-ph/0405447}{{\tt arXiv:astro-ph/0405447}}.
\bibitem[{{Porquet} et~al.(2004){Porquet}, {Reeves}, {O'Brien} and
  {Brinkmann}}]{Porquet+etal+2004}
\bibinfo{author}{{Porquet}, D.}, \bibinfo{author}{{Reeves}, J.N.},
  \bibinfo{author}{{O'Brien}, P.}, \bibinfo{author}{{Brinkmann}, W.},
  \bibinfo{year}{2004}.
\newblock \bibinfo{title}{{XMM-Newton EPIC observations of 21 low-redshift PG
  quasars}}.
\newblock \bibinfo{journal}{\aap} \bibinfo{volume}{422},
  \bibinfo{pages}{85--95}.
\newblock \DOIprefix\doi{10.1051/0004-6361:20047108},
  \href{http://arxiv.org/abs/astro-ph/0404385}{{\tt arXiv:astro-ph/0404385}}.
\bibitem[{{Ramos Almeida} et~al.(2016){Ramos Almeida}, {Mart{\'\i}nez
  Gonz{\'a}lez}, {Asensio Ramos}, {Acosta-Pulido}, {H{\"o}nig},
  {Alonso-Herrero}, {Tadhunter} and
  {Gonz{\'a}lez-Mart{\'\i}n}}]{RamosAlmeida+etal+2016}
\bibinfo{author}{{Ramos Almeida}, C.}, \bibinfo{author}{{Mart{\'\i}nez
  Gonz{\'a}lez}, M.J.}, \bibinfo{author}{{Asensio Ramos}, A.},
  \bibinfo{author}{{Acosta-Pulido}, J.A.}, \bibinfo{author}{{H{\"o}nig}, S.F.},
  \bibinfo{author}{{Alonso-Herrero}, A.}, \bibinfo{author}{{Tadhunter}, C.N.},
  \bibinfo{author}{{Gonz{\'a}lez-Mart{\'\i}n}, O.}, \bibinfo{year}{2016}.
\newblock \bibinfo{title}{{Upholding the unified model for active galactic
  nuclei: VLT/FORS2 spectropolarimetry of Seyfert 2 galaxies}}.
\newblock \bibinfo{journal}{\mnras} \bibinfo{volume}{461},
  \bibinfo{pages}{1387--1403}.
\newblock \DOIprefix\doi{10.1093/mnras/stw1388},
  \href{http://arxiv.org/abs/1606.02204}{{\tt arXiv:1606.02204}}.
\bibitem[{{Ramos Almeida} and {Ricci}(2017)}]{RamosAlmeida+etal+2017}
\bibinfo{author}{{Ramos Almeida}, C.}, \bibinfo{author}{{Ricci}, C.},
  \bibinfo{year}{2017}.
\newblock \bibinfo{title}{{Nuclear obscuration in active galactic nuclei}}.
\newblock \bibinfo{journal}{Nature Astronomy} \bibinfo{volume}{1},
  \bibinfo{pages}{679--689}.
\newblock \DOIprefix\doi{10.1038/s41550-017-0232-z},
  \href{http://arxiv.org/abs/1709.00019}{{\tt arXiv:1709.00019}}.
\bibitem[{{Reeves} and {Turner}(2000)}]{Reeves+etal+2000}
\bibinfo{author}{{Reeves}, J.N.}, \bibinfo{author}{{Turner}, M.J.L.},
  \bibinfo{year}{2000}.
\newblock \bibinfo{title}{{X-ray spectra of a large sample of quasars with
  ASCA}}.
\newblock \bibinfo{journal}{\mnras} \bibinfo{volume}{316},
  \bibinfo{pages}{234--248}.
\newblock \DOIprefix\doi{10.1046/j.1365-8711.2000.03510.x},
  \href{http://arxiv.org/abs/astro-ph/0003080}{{\tt arXiv:astro-ph/0003080}}.
\bibitem[{{Ricci} et~al.(2013){Ricci}, {Paltani}, {Ueda} and
  {Awaki}}]{Ricci+etal+2013}
\bibinfo{author}{{Ricci}, C.}, \bibinfo{author}{{Paltani}, S.},
  \bibinfo{author}{{Ueda}, Y.}, \bibinfo{author}{{Awaki}, H.},
  \bibinfo{year}{2013}.
\newblock \bibinfo{title}{{On the role of the {\ensuremath{\Gamma}} -
  {\ensuremath{\lambda}}$_{Edd}$ relation on the X-ray Baldwin effect in active
  galactic nuclei}}.
\newblock \bibinfo{journal}{\mnras} \bibinfo{volume}{435},
  \bibinfo{pages}{1840--1851}.
\newblock \DOIprefix\doi{10.1093/mnras/stt1326},
  \href{http://arxiv.org/abs/1307.4507}{{\tt arXiv:1307.4507}}.
\bibitem[{{Risaliti} et~al.(2005){Risaliti}, {Elvis}, {Fabbiano}, {Baldi} and
  {Zezas}}]{Risaliti+etal+2005}
\bibinfo{author}{{Risaliti}, G.}, \bibinfo{author}{{Elvis}, M.},
  \bibinfo{author}{{Fabbiano}, G.}, \bibinfo{author}{{Baldi}, A.},
  \bibinfo{author}{{Zezas}, A.}, \bibinfo{year}{2005}.
\newblock \bibinfo{title}{{Rapid Compton-thick/Compton-thin Transitions in the
  Seyfert 2 Galaxy NGC 1365}}.
\newblock \bibinfo{journal}{\apjl} \bibinfo{volume}{623},
  \bibinfo{pages}{L93--L96}.
\newblock \DOIprefix\doi{10.1086/430252},
  \href{http://arxiv.org/abs/astro-ph/0503351}{{\tt arXiv:astro-ph/0503351}}.
\bibitem[{{Risaliti} et~al.(2011){Risaliti}, {Nardini}, {Salvati}, {Elvis},
  {Fabbiano}, {Maiolino}, {Pietrini} and
  {Torricelli-Ciamponi}}]{Risaliti+etal+2011}
\bibinfo{author}{{Risaliti}, G.}, \bibinfo{author}{{Nardini}, E.},
  \bibinfo{author}{{Salvati}, M.}, \bibinfo{author}{{Elvis}, M.},
  \bibinfo{author}{{Fabbiano}, G.}, \bibinfo{author}{{Maiolino}, R.},
  \bibinfo{author}{{Pietrini}, P.}, \bibinfo{author}{{Torricelli-Ciamponi},
  G.}, \bibinfo{year}{2011}.
\newblock \bibinfo{title}{{X-ray absorption by broad-line region clouds in Mrk
  766}}.
\newblock \bibinfo{journal}{\mnras} \bibinfo{volume}{410},
  \bibinfo{pages}{1027--1035}.
\newblock \DOIprefix\doi{10.1111/j.1365-2966.2010.17503.x},
  \href{http://arxiv.org/abs/1008.5067}{{\tt arXiv:1008.5067}}.
\bibitem[{{Savi{\'c}} et~al.(2020){Savi{\'c}}, {Popovi{\'c}}, {Shablovinskaya}
  and {Afanasiev}}]{savic+etal+2020}
\bibinfo{author}{{Savi{\'c}}, D.}, \bibinfo{author}{{Popovi{\'c}}, L.{\v{C}}.},
  \bibinfo{author}{{Shablovinskaya}, E.}, \bibinfo{author}{{Afanasiev}, V.L.},
  \bibinfo{year}{2020}.
\newblock \bibinfo{title}{{Estimating supermassive black hole masses in active
  galactic nuclei using polarization of broad Mg II, H {\ensuremath{\alpha}},
  and H {\ensuremath{\beta}} lines}}.
\newblock \bibinfo{journal}{\mnras} \bibinfo{volume}{497},
  \bibinfo{pages}{3047--3054}.
\newblock \DOIprefix\doi{10.1093/mnras/staa2039},
  \href{http://arxiv.org/abs/2007.11475}{{\tt arXiv:2007.11475}}.
\bibitem[{{Schlegel} et~al.(1998){Schlegel}, {Finkbeiner} and
  {Davis}}]{Schlegel+etal+1998}
\bibinfo{author}{{Schlegel}, D.J.}, \bibinfo{author}{{Finkbeiner}, D.P.},
  \bibinfo{author}{{Davis}, M.}, \bibinfo{year}{1998}.
\newblock \bibinfo{title}{{Maps of Dust Infrared Emission for Use in Estimation
  of Reddening and Cosmic Microwave Background Radiation Foregrounds}}.
\newblock \bibinfo{journal}{\apj} \bibinfo{volume}{500},
  \bibinfo{pages}{525--553}.
\newblock \DOIprefix\doi{10.1086/305772},
  \href{http://arxiv.org/abs/astro-ph/9710327}{{\tt arXiv:astro-ph/9710327}}.
\bibitem[{{Scott} et~al.(2011){Scott}, {Stewart}, {Mateos}, {Alexander},
  {Hutton} and {Ward}}]{Scott+etal+2011}
\bibinfo{author}{{Scott}, A.E.}, \bibinfo{author}{{Stewart}, G.C.},
  \bibinfo{author}{{Mateos}, S.}, \bibinfo{author}{{Alexander}, D.M.},
  \bibinfo{author}{{Hutton}, S.}, \bibinfo{author}{{Ward}, M.J.},
  \bibinfo{year}{2011}.
\newblock \bibinfo{title}{{New constraints on the X-ray spectral properties of
  type 1 active galactic nuclei}}.
\newblock \bibinfo{journal}{\mnras} \bibinfo{volume}{417},
  \bibinfo{pages}{992--1012}.
\newblock \DOIprefix\doi{10.1111/j.1365-2966.2011.19325.x},
  \href{http://arxiv.org/abs/1106.4904}{{\tt arXiv:1106.4904}}.
\bibitem[{{Shapovalova} et~al.(2012){Shapovalova}, {Popovi{\'c}}, {Burenkov},
  {Chavushyan}, {Ili{\'c}}, {Kova{\v{c}}evi{\'c}}, {Kollatschny},
  {Kova{\v{c}}evi{\'c}}, {Bochkarev}, {Valdes}, {Torrealba},
  {Le{\'o}n-Tavares}, {Mercado}, {Ben{\'\i}tez}, {Carrasco}, {Dultzin} and {de
  la Fuente}}]{Shapovalova+etal+2012}
\bibinfo{author}{{Shapovalova}, A.I.}, \bibinfo{author}{{Popovi{\'c}},
  L.{\v{C}}.}, \bibinfo{author}{{Burenkov}, A.N.},
  \bibinfo{author}{{Chavushyan}, V.H.}, \bibinfo{author}{{Ili{\'c}}, D.},
  \bibinfo{author}{{Kova{\v{c}}evi{\'c}}, A.}, \bibinfo{author}{{Kollatschny},
  W.}, \bibinfo{author}{{Kova{\v{c}}evi{\'c}}, J.},
  \bibinfo{author}{{Bochkarev}, N.G.}, \bibinfo{author}{{Valdes}, J.R.},
  \bibinfo{author}{{Torrealba}, J.}, \bibinfo{author}{{Le{\'o}n-Tavares}, J.},
  \bibinfo{author}{{Mercado}, A.}, \bibinfo{author}{{Ben{\'\i}tez}, E.},
  \bibinfo{author}{{Carrasco}, L.}, \bibinfo{author}{{Dultzin}, D.},
  \bibinfo{author}{{de la Fuente}, E.}, \bibinfo{year}{2012}.
\newblock \bibinfo{title}{{Spectral Optical Monitoring of the Narrow-line
  Seyfert 1 Galaxy Ark 564}}.
\newblock \bibinfo{journal}{\apjs} \bibinfo{volume}{202}, \bibinfo{pages}{10}.
\newblock \DOIprefix\doi{10.1088/0067-0049/202/1/10},
  \href{http://arxiv.org/abs/1207.1782}{{\tt arXiv:1207.1782}}.
\bibitem[{{Shehata} et~al.(2021){Shehata}, {Misra}, {Osman}, {Shalabiea} and
  {Hayman}}]{Shehata+etal+2021}
\bibinfo{author}{{Shehata}, S.M.}, \bibinfo{author}{{Misra}, R.},
  \bibinfo{author}{{Osman}, A.M.I.}, \bibinfo{author}{{Shalabiea}, O.M.},
  \bibinfo{author}{{Hayman}, Z.M.}, \bibinfo{year}{2021}.
\newblock \bibinfo{title}{{Redshift evolution of X-ray spectral index of
  quasars observed by XMM-NEWTON/SDSS}}.
\newblock \bibinfo{journal}{Journal of High Energy Astrophysics}
  \bibinfo{volume}{31}, \bibinfo{pages}{37--43}.
\newblock \DOIprefix\doi{10.1016/j.jheap.2021.04.003}.
\bibitem[{{Shemmer} et~al.(2006){Shemmer}, {Brandt}, {Netzer}, {Maiolino} and
  {Kaspi}}]{Shemmer+etal+2006}
\bibinfo{author}{{Shemmer}, O.}, \bibinfo{author}{{Brandt}, W.N.},
  \bibinfo{author}{{Netzer}, H.}, \bibinfo{author}{{Maiolino}, R.},
  \bibinfo{author}{{Kaspi}, S.}, \bibinfo{year}{2006}.
\newblock \bibinfo{title}{{The Hard X-Ray Spectral Slope as an Accretion Rate
  Indicator in Radio-quiet Active Galactic Nuclei}}.
\newblock \bibinfo{journal}{\apjl} \bibinfo{volume}{646},
  \bibinfo{pages}{L29--L32}.
\newblock \DOIprefix\doi{10.1086/506911},
  \href{http://arxiv.org/abs/astro-ph/0606389}{{\tt arXiv:astro-ph/0606389}}.
\bibitem[{{Shemmer} et~al.(2008){Shemmer}, {Brandt}, {Netzer}, {Maiolino} and
  {Kaspi}}]{Shemmer+etal+2008}
\bibinfo{author}{{Shemmer}, O.}, \bibinfo{author}{{Brandt}, W.N.},
  \bibinfo{author}{{Netzer}, H.}, \bibinfo{author}{{Maiolino}, R.},
  \bibinfo{author}{{Kaspi}, S.}, \bibinfo{year}{2008}.
\newblock \bibinfo{title}{{The Hard X-Ray Spectrum as a Probe for Black Hole
  Growth in Radio-Quiet Active Galactic Nuclei}}.
\newblock \bibinfo{journal}{\apj} \bibinfo{volume}{682},
  \bibinfo{pages}{81--93}.
\newblock \DOIprefix\doi{10.1086/588776},
  \href{http://arxiv.org/abs/0804.0803}{{\tt arXiv:0804.0803}}.
\bibitem[{{Shemmer} et~al.(2005){Shemmer}, {Brandt}, {Vignali}, {Schneider},
  {Fan}, {Richards} and {Strauss}}]{Shemmer+etal+2005}
\bibinfo{author}{{Shemmer}, O.}, \bibinfo{author}{{Brandt}, W.N.},
  \bibinfo{author}{{Vignali}, C.}, \bibinfo{author}{{Schneider}, D.P.},
  \bibinfo{author}{{Fan}, X.}, \bibinfo{author}{{Richards}, G.T.},
  \bibinfo{author}{{Strauss}, M.A.}, \bibinfo{year}{2005}.
\newblock \bibinfo{title}{{The X-Ray Spectral Properties and Variability of
  Luminous High-Redshift Active Galactic Nuclei}}.
\newblock \bibinfo{journal}{\apj} \bibinfo{volume}{630},
  \bibinfo{pages}{729--739}.
\newblock \DOIprefix\doi{10.1086/432050},
  \href{http://arxiv.org/abs/astro-ph/0505482}{{\tt arXiv:astro-ph/0505482}}.
\bibitem[{{Shen} et~al.(2011){Shen}, {Richards}, {Strauss}, {Hall},
  {Schneider}, {Snedden}, {Bizyaev}, {Brewington}, {Malanushenko},
  {Malanushenko}, {Oravetz}, {Pan} and {Simmons}}]{Shen+etal+2011}
\bibinfo{author}{{Shen}, Y.}, \bibinfo{author}{{Richards}, G.T.},
  \bibinfo{author}{{Strauss}, M.A.}, \bibinfo{author}{{Hall}, P.B.},
  \bibinfo{author}{{Schneider}, D.P.}, \bibinfo{author}{{Snedden}, S.},
  \bibinfo{author}{{Bizyaev}, D.}, \bibinfo{author}{{Brewington}, H.},
  \bibinfo{author}{{Malanushenko}, V.}, \bibinfo{author}{{Malanushenko}, E.},
  \bibinfo{author}{{Oravetz}, D.}, \bibinfo{author}{{Pan}, K.},
  \bibinfo{author}{{Simmons}, A.}, \bibinfo{year}{2011}.
\newblock \bibinfo{title}{{A Catalog of Quasar Properties from Sloan Digital
  Sky Survey Data Release 7}}.
\newblock \bibinfo{journal}{\apjs} \bibinfo{volume}{194}, \bibinfo{pages}{45}.
\newblock \DOIprefix\doi{10.1088/0067-0049/194/2/45},
  \href{http://arxiv.org/abs/1006.5178}{{\tt arXiv:1006.5178}}.
\bibitem[{{Sriram} et~al.(2022){Sriram}, {Nour} and {Choi}}]{Sriram+etal+2022}
\bibinfo{author}{{Sriram}, K.}, \bibinfo{author}{{Nour}, D.},
  \bibinfo{author}{{Choi}, C.S.}, \bibinfo{year}{2022}.
\newblock \bibinfo{title}{{Influence of Comptonization region over the ambiance
  of accretion disc in active galactic nucleus}}.
\newblock \bibinfo{journal}{\mnras} \bibinfo{volume}{510},
  \bibinfo{pages}{3222--3235}.
\newblock \DOIprefix\doi{10.1093/mnras/stab3610},
  \href{http://arxiv.org/abs/2112.04180}{{\tt arXiv:2112.04180}}.
\bibitem[{{Trakhtenbrot} et~al.(2017){Trakhtenbrot}, {Ricci}, {Koss},
  {Schawinski}, {Mushotzky}, {Ueda}, {Veilleux}, {Lamperti}, {Oh}, {Treister},
  {Stern}, {Harrison}, {Balokovi{\'c}} and {Gehrels}}]{Trakhtenbrot+etal+2017}
\bibinfo{author}{{Trakhtenbrot}, B.}, \bibinfo{author}{{Ricci}, C.},
  \bibinfo{author}{{Koss}, M.J.}, \bibinfo{author}{{Schawinski}, K.},
  \bibinfo{author}{{Mushotzky}, R.}, \bibinfo{author}{{Ueda}, Y.},
  \bibinfo{author}{{Veilleux}, S.}, \bibinfo{author}{{Lamperti}, I.},
  \bibinfo{author}{{Oh}, K.}, \bibinfo{author}{{Treister}, E.},
  \bibinfo{author}{{Stern}, D.}, \bibinfo{author}{{Harrison}, F.},
  \bibinfo{author}{{Balokovi{\'c}}, M.}, \bibinfo{author}{{Gehrels}, N.},
  \bibinfo{year}{2017}.
\newblock \bibinfo{title}{{BAT AGN Spectroscopic Survey (BASS) - VI. The
  {\ensuremath{\Gamma}}$_{X}$-L/L$_{Edd}$ relation}}.
\newblock \bibinfo{journal}{\mnras} \bibinfo{volume}{470},
  \bibinfo{pages}{800--814}.
\newblock \DOIprefix\doi{10.1093/mnras/stx1117},
  \href{http://arxiv.org/abs/1705.01550}{{\tt arXiv:1705.01550}}.
\bibitem[{{Urry} and {Padovani}(1995)}]{Urry+etal+1995}
\bibinfo{author}{{Urry}, C.M.}, \bibinfo{author}{{Padovani}, P.},
  \bibinfo{year}{1995}.
\newblock \bibinfo{title}{{Unified Schemes for Radio-Loud Active Galactic
  Nuclei}}.
\newblock \bibinfo{journal}{\pasp} \bibinfo{volume}{107}, \bibinfo{pages}{803}.
\newblock \DOIprefix\doi{10.1086/133630},
  \href{http://arxiv.org/abs/astro-ph/9506063}{{\tt arXiv:astro-ph/9506063}}.
\bibitem[{{Vanden Berk} et~al.(2006){Vanden Berk}, {Shen}, {Yip}, {Schneider},
  {Connolly}, {Burton}, {Jester}, {Hall}, {Szalay} and
  {Brinkmann}}]{VandenBerk+etal+2006}
\bibinfo{author}{{Vanden Berk}, D.E.}, \bibinfo{author}{{Shen}, J.},
  \bibinfo{author}{{Yip}, C.W.}, \bibinfo{author}{{Schneider}, D.P.},
  \bibinfo{author}{{Connolly}, A.J.}, \bibinfo{author}{{Burton}, R.E.},
  \bibinfo{author}{{Jester}, S.}, \bibinfo{author}{{Hall}, P.B.},
  \bibinfo{author}{{Szalay}, A.S.}, \bibinfo{author}{{Brinkmann}, J.},
  \bibinfo{year}{2006}.
\newblock \bibinfo{title}{{Spectral Decomposition of Broad-Line AGNs and Host
  Galaxies}}.
\newblock \bibinfo{journal}{\aj} \bibinfo{volume}{131},
  \bibinfo{pages}{84--99}.
\newblock \DOIprefix\doi{10.1086/497973},
  \href{http://arxiv.org/abs/astro-ph/0509332}{{\tt arXiv:astro-ph/0509332}}.
\bibitem[{{Vasudevan} and {Fabian}(2009)}]{Vasudevan+etal+2009}
\bibinfo{author}{{Vasudevan}, R.V.}, \bibinfo{author}{{Fabian}, A.C.},
  \bibinfo{year}{2009}.
\newblock \bibinfo{title}{{Simultaneous X-ray/optical/UV snapshots of active
  galactic nuclei from XMM-Newton: spectral energy distributions for the
  reverberation mapped sample}}.
\newblock \bibinfo{journal}{\mnras} \bibinfo{volume}{392},
  \bibinfo{pages}{1124--1140}.
\newblock \DOIprefix\doi{10.1111/j.1365-2966.2008.14108.x},
  \href{http://arxiv.org/abs/0810.3777}{{\tt arXiv:0810.3777}}.
\bibitem[{{Winter} et~al.(2012){Winter}, {Veilleux}, {McKernan} and
  {Kallman}}]{Winter+etal+2012}
\bibinfo{author}{{Winter}, L.M.}, \bibinfo{author}{{Veilleux}, S.},
  \bibinfo{author}{{McKernan}, B.}, \bibinfo{author}{{Kallman}, T.R.},
  \bibinfo{year}{2012}.
\newblock \bibinfo{title}{{The Swift Burst Alert Telescope Detected Seyfert 1
  Galaxies: X-Ray Broadband Properties and Warm Absorbers}}.
\newblock \bibinfo{journal}{\apj} \bibinfo{volume}{745}, \bibinfo{pages}{107}.
\newblock \DOIprefix\doi{10.1088/0004-637X/745/2/107},
  \href{http://arxiv.org/abs/1112.0540}{{\tt arXiv:1112.0540}}.
\bibitem[{{Xie} et~al.(2017){Xie}, {Yuan} and {Ho}}]{Xie+rtal+2017}
\bibinfo{author}{{Xie}, F.G.}, \bibinfo{author}{{Yuan}, F.},
  \bibinfo{author}{{Ho}, L.C.}, \bibinfo{year}{2017}.
\newblock \bibinfo{title}{{Radiative Heating in the Kinetic Mode of AGN
  Feedback}}.
\newblock \bibinfo{journal}{\apj} \bibinfo{volume}{844}, \bibinfo{pages}{42}.
\newblock \DOIprefix\doi{10.3847/1538-4357/aa7950},
  \href{http://arxiv.org/abs/1705.05037}{{\tt arXiv:1705.05037}}.
\bibitem[{{XRISM Science Team}(2022)}]{XRISM+etal+2022}
\bibinfo{author}{{XRISM Science Team}}, \bibinfo{year}{2022}.
\newblock \bibinfo{title}{{XRISM Quick Reference}}.
\newblock \bibinfo{journal}{arXiv e-prints} ,
  \bibinfo{pages}{arXiv:2202.05399}\href{http://arxiv.org/abs/2202.05399}{{\tt
  arXiv:2202.05399}}.
\bibitem[{{Yip} et~al.(2004a){Yip}, {Connolly}, {Szalay}, {Budav{\'a}ri},
  {SubbaRao}, {Frieman}, {Nichol}, {Hopkins}, {York}, {Okamura}, {Brinkmann},
  {Csabai}, {Thakar}, {Fukugita} and {Ivezi{\'c}}}]{Yip+etal+2004a}
\bibinfo{author}{{Yip}, C.W.}, \bibinfo{author}{{Connolly}, A.J.},
  \bibinfo{author}{{Szalay}, A.S.}, \bibinfo{author}{{Budav{\'a}ri}, T.},
  \bibinfo{author}{{SubbaRao}, M.}, \bibinfo{author}{{Frieman}, J.A.},
  \bibinfo{author}{{Nichol}, R.C.}, \bibinfo{author}{{Hopkins}, A.M.},
  \bibinfo{author}{{York}, D.G.}, \bibinfo{author}{{Okamura}, S.},
  \bibinfo{author}{{Brinkmann}, J.}, \bibinfo{author}{{Csabai}, I.},
  \bibinfo{author}{{Thakar}, A.R.}, \bibinfo{author}{{Fukugita}, M.},
  \bibinfo{author}{{Ivezi{\'c}}, {\v{Z}}.}, \bibinfo{year}{2004}a.
\newblock \bibinfo{title}{{Distributions of Galaxy Spectral Types in the Sloan
  Digital Sky Survey}}.
\newblock \bibinfo{journal}{\aj} \bibinfo{volume}{128},
  \bibinfo{pages}{585--609}.
\newblock \DOIprefix\doi{10.1086/422429},
  \href{http://arxiv.org/abs/astro-ph/0407061}{{\tt arXiv:astro-ph/0407061}}.
\bibitem[{{Yip} et~al.(2004b){Yip}, {Connolly}, {Vanden Berk}, {Ma}, {Frieman},
  {SubbaRao}, {Szalay}, {Richards}, {Hall}, {Schneider}, {Hopkins}, {Trump} and
  {Brinkmann}}]{Yip+etal+2004b}
\bibinfo{author}{{Yip}, C.W.}, \bibinfo{author}{{Connolly}, A.J.},
  \bibinfo{author}{{Vanden Berk}, D.E.}, \bibinfo{author}{{Ma}, Z.},
  \bibinfo{author}{{Frieman}, J.A.}, \bibinfo{author}{{SubbaRao}, M.},
  \bibinfo{author}{{Szalay}, A.S.}, \bibinfo{author}{{Richards}, G.T.},
  \bibinfo{author}{{Hall}, P.B.}, \bibinfo{author}{{Schneider}, D.P.},
  \bibinfo{author}{{Hopkins}, A.M.}, \bibinfo{author}{{Trump}, J.},
  \bibinfo{author}{{Brinkmann}, J.}, \bibinfo{year}{2004}b.
\newblock \bibinfo{title}{{Spectral Classification of Quasars in the Sloan
  Digital Sky Survey: Eigenspectra, Redshift, and Luminosity Effects}}.
\newblock \bibinfo{journal}{\aj} \bibinfo{volume}{128},
  \bibinfo{pages}{2603--2630}.
\newblock \DOIprefix\doi{10.1086/425626},
  \href{http://arxiv.org/abs/astro-ph/0408578}{{\tt arXiv:astro-ph/0408578}}.
\bibitem[{{Young} et~al.(2009){Young}, {Elvis} and
  {Risaliti}}]{Young+etal+2009}
\bibinfo{author}{{Young}, M.}, \bibinfo{author}{{Elvis}, M.},
  \bibinfo{author}{{Risaliti}, G.}, \bibinfo{year}{2009}.
\newblock \bibinfo{title}{{The Fifth Data Release Sloan Digital Sky
  Survey/XMM-Newton Quasar Survey}}.
\newblock \bibinfo{journal}{\apjs} \bibinfo{volume}{183},
  \bibinfo{pages}{17--32}.
\newblock \DOIprefix\doi{10.1088/0067-0049/183/1/17},
  \href{http://arxiv.org/abs/0905.0496}{{\tt arXiv:0905.0496}}.
\bibitem[{{Yuan} and {Narayan}(2014)}]{Yuan+etal+2014}
\bibinfo{author}{{Yuan}, F.}, \bibinfo{author}{{Narayan}, R.},
  \bibinfo{year}{2014}.
\newblock \bibinfo{title}{{Hot Accretion Flows Around Black Holes}}.
\newblock \bibinfo{journal}{\araa} \bibinfo{volume}{52},
  \bibinfo{pages}{529--588}.
\newblock \DOIprefix\doi{10.1146/annurev-astro-082812-141003},
  \href{http://arxiv.org/abs/1401.0586}{{\tt arXiv:1401.0586}}.
\bibitem[{{Yuan} et~al.(1998){Yuan}, {Brinkmann}, {Siebert} and
  {Voges}}]{Yuan+etal+1998}
\bibinfo{author}{{Yuan}, W.}, \bibinfo{author}{{Brinkmann}, W.},
  \bibinfo{author}{{Siebert}, J.}, \bibinfo{author}{{Voges}, W.},
  \bibinfo{year}{1998}.
\newblock \bibinfo{title}{{Broad band energy distribution of ROSAT detected
  quasars. II. Radio-quiet objects}}.
\newblock \bibinfo{journal}{\aap} \bibinfo{volume}{330},
  \bibinfo{pages}{108--122}.
\newblock \href{http://arxiv.org/abs/astro-ph/9805015}{{\tt
  arXiv:astro-ph/9805015}}.
\bibitem[{{Zhou} and {Zhao}(2010)}]{Zhou+etal+2010}
\bibinfo{author}{{Zhou}, X.L.}, \bibinfo{author}{{Zhao}, Y.H.},
  \bibinfo{year}{2010}.
\newblock \bibinfo{title}{{Hard X-ray Photon Index as an Indicator of
  Bolometric Correction in Active Galactic Nuclei}}.
\newblock \bibinfo{journal}{\apjl} \bibinfo{volume}{720},
  \bibinfo{pages}{L206--L210}.
\newblock \DOIprefix\doi{10.1088/2041-8205/720/2/L206},
  \href{http://arxiv.org/abs/1008.2532}{{\tt arXiv:1008.2532}}.
\bibitem[{{Zoghbi} et~al.(2012){Zoghbi}, {Fabian}, {Reynolds} and
  {Cackett}}]{Zoghbi+etal+2012}
\bibinfo{author}{{Zoghbi}, A.}, \bibinfo{author}{{Fabian}, A.C.},
  \bibinfo{author}{{Reynolds}, C.S.}, \bibinfo{author}{{Cackett}, E.M.},
  \bibinfo{year}{2012}.
\newblock \bibinfo{title}{{Relativistic iron K X-ray reverberation in NGC
  4151}}.
\newblock \bibinfo{journal}{\mnras} \bibinfo{volume}{422},
  \bibinfo{pages}{129--134}.
\newblock \DOIprefix\doi{10.1111/j.1365-2966.2012.20587.x},
  \href{http://arxiv.org/abs/1112.1717}{{\tt arXiv:1112.1717}}.

\end{thebibliography}

\bio{}
\endbio

\bio{}
\endbio

\end{document}